\documentclass[preprint]{aastex}

\newcommand{\ms}{\ensuremath{M_{\odot}}}
\newcommand{\mdot}{\ensuremath{\dot{M}}}
\newcommand{\fup}{\ensuremath{f_{u}}}
\newcommand{\fdown}{\ensuremath{f_{d}}}

\shorttitle{Cool bottom processes on the AGB}
\begin{document}

\received{}
\accepted{}

\title{Cool bottom processes on the thermally-pulsing AGB and the isotopic
composition of circumstellar dust grains} %

\author{Kenneth M. Nollett} \affil{MC 130-33, California Institute of
Technology, Pasadena, CA 91125} \email{nollett@caltech.edu}

\author{M. Busso}
\affil{Department of Physics, University of Perugia, Via Pascoli, 06100
Perugia,
Italy}
\email{busso@fisica.unipg.it}

\and

\author{G. J. Wasserburg}
\affil{The Lunatic Asylum, Division of Geological and Planetary
Sciences,\\ California Institute of Technology, Pasadena,
CA~91125, USA}
\email{isotopes@gps.caltech.edu}

\begin{abstract}

We examine the effects of cool bottom processing (CBP) on the isotopic
ratios $^{18}$O/$^{16}$O, $^{17}$O/$^{16}$O, $^{14}$N/$^{15}$N,
$^{26}$Al/$^{27}$Al, C/O, and N/O in the convective envelope during
the thermally-pulsing asymptotic giant branch (TP-AGB) phase of
evolution in a 1.5\ms\ initial-mass star of solar initial composition.
We use a parametric model which treats extra mixing by introducing
mass flow between the convective envelope and the underlying radiative
zone.  The parameters of this model are the mass circulation rate
(\mdot) and the maximum temperature ($T_P$) experienced by the
circulating material.  The effects of nuclear reactions in the flowing
matter were calculated using a set of static structures of the
radiative zone selected from particular times in a complete stellar
evolution calculation.  The compositions of the flowing material were
obtained and the resulting changes in the envelope determined.  No
major shifts in the star's energy budget occur from the imposed CBP if
$\log T_P< 7.73$.  Using structures from several times on the TP-AGB,
it was found that the results for all species except $^{26}$Al were
essentially independent of the time chosen if $\log T_P > 7.6$.
Abundant $^{26}$Al was produced by CBP for $\log T_P>7.65$.  While
$^{26}$Al/$^{27}$Al depends on $T_P$, the other isotopic ratios depend
dominantly on the circulation rate.  The relationship is shown between
models of CBP as parameterized by a diffusion formalism within the
stellar evolution model and those using the mass-flow formalism
employed here.  They are shown to be effectively equivalent.  In
general, the CBP treatment readily permits calculation of envelope
compositions as affected by different degrees of extra mixing, based
on stellar structures computed by normal stellar evolution
models. Using these results, the isotopic ratios under conditions of
C/O$<1$ and C/O$>1$ are compared with the data on circumstellar dust
grains.  It is found that the $^{18}$O/$^{16}$O, $^{17}$O/$^{16}$O,
and $^{26}$Al/$^{27}$Al observed for oxide grains formed at C/O$<1$
are reasonably well-understood.  However, the $^{15}$N/$^{14}$N,
$^{12}$C/$^{13}$C, and $^{26}$Al/$^{27}$Al in carbide grains (C/O$>1$)
require that many of their stellar sources must have had
$^{14}$N/$^{15}$N at least a factor of 4 lower than the solar value.
This allows a self-consistent description of all these isotopes in
most SiC grains.  The rare grains with $^{12}$C/$^{13}$C$<10$ cannot
be produced by any red-giant or AGB source, nor are they reconcileable
with novae sources.

\end{abstract}


\section{Introduction}
\label{sec:intro}

The purpose of this study is to explore the effects of an extra
circulation process on the $^{26}$Al/$^{27}$Al, $^{17}$O/$^{16}$O,
$^{18}$O/$^{16}$O, and $^{13}$C/$^{12}$C ratios in the envelope of an
AGB star of solar metallicity. By an extra circulation process we mean
the transport of matter from the fully convective envelope into the
underlying radiative region, down to the outer parts of the H-burning
shell.  The actual physical-dynamical basis for such penetration is
not known.  However, the requirement that some such mechanism must
sometime occur has been amply demonstrated by observations on RGB
stars and in studies of dust grains which formed in circumstellar
envelopes (e.g. Huss et al. 1994; Boothroyd, Sackmann \& Wasserburg
1994; Harris, Lambert \& Smith 1985; Kahane et al. 1992).  The
isotopic effects of such extra transport on the abundances depend only
on the degree of nuclear processing. This is directly related to the
nuclear reaction rates, the rate of mass transfer and the depth to
which the extra mixing mechanism penetrates. A detailed physical
fluid-dynamical mechanism can not significantly affect the resulting
abundances beyond determining the rate and depth.  It is clear that
any extra mixing mechanism that does not fractionate one set of
elements relative to the other elements involved in the nuclear
reactions will not change the resulting products that are added to the
envelope from what is found here.  The issue of how small amounts of
extra mixing will change the stellar evolution is not known but will
be commented on. The results, presented here with inter-relationships
between the different isotopic ratios, are of use in discussing and
understanding isotopic results obtained by extensive studies of
recovered circumstellar dust grains.  Insofar as a relationship may be
found between the isotopic ratios listed above and the ratios of major
elements in the AGB envelope (C/O and N/O), this may lead to
understanding the chemical-physical origins of the grains and the
nature of the stars around which they formed. Such results may also be
of value in understanding the compositions of planetary nebulae.

  It has been proposed that many short-lived nuclei present in the
early solar system (ESS) were the products of an AGB star that
injected freshly-synthesized nuclei into the interstellar medium (ISM)
from which the solar system formed (Wasserburg et al. 1994). Because
AGB stars are the site of $s$-process nucleosynthesis, there is a
rather thorough understanding of the yields of a large number of
nuclides in their convective envelopes (cf. Busso, Gallino \&
Wasserburg 1999, hereafter BGW99).  Among these nuclides, $^{26}$Al
plays a critical role in evaluating the hypothesis that an AGB star
supplied short-lived nuclei to the early solar system.  While it is
clear that $^{26}$Al should be produced in AGB stars by the reaction
$^{25}$Mg($p,\gamma$)$^{26}$Al (Forestini, Arnould \& Paulus 1991;
Mowlavi \& Meynet 2000; Lattanzio, Forestini \& Charbonnel 2000), the
quantitative $^{26}$Al yield, taking into account the complexities of
destruction of this nuclide, remains very uncertain.  As shown by
Wasserburg et al. (1994) and BGW99, $^{26}$Al is produced efficiently
in the hydrogen shell, but most ($\sim 90 \%$) of this is mixed into
the He shell at the onset of intermittent He shell burning.  The mixed
material is then exposed to neutrons in the convective (He-burning)
region formed during the He-burning pulse, where 80\% to 100\% of its
$^{26}$Al content is consumed by the reaction $^{26}{\rm
Al}(n,p)^{26}{\rm Mg}$. The neutrons that destroy $^{26}$Al are
provided by the process $^{22}{\rm Ne}(\alpha,n)^{25}{\rm Mg}$,
activated marginally in the relatively high temperatures ($T \geq
3\times 10^{8}$ K) of the thermal pulse.  A value of
$^{26}$Al/$^{27}$Al$\sim 10^{-3}$ ($\equiv$ number ratio) results in
the envelope after mixing and is much less than the amount initially
produced in the hydrogen shell.  This is a factor of a few less than
required if a single AGB star produced the $^{26}$Al and this material
was then diluted to provide the amount found in the early solar system
($^{26}$Al/$^{27}$Al$\sim 5\times 10^{-5}$; Lee, Papanastassiou \&
Wasserburg 1977; Wasserburg et al. 1994, 1995). Although the
discrepancy might be explained by a relatively small change in the
fraction of $^{26}$Al destroyed (i.e., a small variation in the
$^{26}{\rm Al}(n,p)^{26}{\rm Mg}$ rate), the estimated $^{26}$Al yield
is at best in marginal agreement with the yield required to provide
the solar inventory. If the true yield is lower than $^{26}{\rm
Al}/^{27}{\rm Al} \sim 4\times 10^{-3}$, then an AGB source for solar
$^{26}$Al is not reasonable. There is at present no way to improve the
estimated $^{26}$Al yield.  Mowlavi \& Meynet (2000) recently carried
out an extensive analysis of $^{26}$Al production in AGB stars.  They
explored the effects of recently-revised reaction rates, and computed
the yields expected at varying metallicities and initial masses.  The
results of this work agree with the above expectations: normal,
low-mass AGB stars were found to be probable sources of $^{26}$Al
abundances observed in most (not all) SiC grains, with
$^{26}$Al/$^{27}$Al values in the same range discussed above. Mowlavi
\& Meynet also found more efficient production of $^{26}$Al in
intermediate mass stars (hereafter IMS) experiencing hot bottom
burning (HBB).  This process was studied extensively by Frost et
al. (1998), Lattanzio \& Forestini (1999), and Lattanzio et
al. (2000).  However, uncertainties in the final fates of these stars
(mainly in the number of thermal pulses experienced) prevent a
quantitative estimate of their possible contributions to the ISM and
to the Galactic inventory.

A direct challenge to standard AGB models arises from the discovery of
circumstellar oxide grains recovered from meteorites, on which rather
precise isotopic ratios are measured.  Many of these grains reflect
$^{26}$Al/$^{27}$Al at their formation far above $10^{-3}$.  Some
oxide grains also have $^{18}$O/$^{16}$O about 30\% less than the
solar value and $^{17}$O/$^{16}$O much greater than the solar value.
These oxygen data were interpreted to reflect stellar first dredge-up
(e.g., Huss et al. 1994, Boothroyd, Sackmann \& Wasserburg 1994).
Studies by Nittler et al. (1994,1997) and Choi, Wasserburg \& Huss
(1999) found some oxide grains with very low or essentially no
$^{18}$O.  It was shown that such low $^{18}$O/$^{16}$O could result
from HBB (Boothroyd, Sackmann, \& Wasserburg, 1995), but only in stars
with masses $4 M_\odot <M< 8 M_{\odot}$.  However, the data on many
oxide grains were found to lie in a region of the $^{17}$O/$^{16}$O
and $^{18}$O/$^{16}$O diagram that is not accessible by HBB.  Direct
observations of oxide molecules in the photospheres of some red giants
had for some time indicated similar oxygen isotopic compositions,
albeit with very large uncertainties (Harris, Lambert \& Smith 1985;
Harris et al. 1987).  Similar modifications of oxygen isotopes were
subsequently confirmed with a higher level of confidence in
circumstellar envelopes (Kahane et al. 1992).

Discrepancies between oxide grain compositions and the compositions
that result from simple dredge-up in low-mass stars were explained by
the hypothesis of extra mixing of material through the zone of
radiative energy transport above the hydrogen shell -- referred to as
``deep mixing'' or ``cool bottom processing'' (CBP).  Such a mixing
process is not treated quantitatively in any stellar model with a
stable and totally non-circulating radiative zone, but it cannot be
ruled out on physical grounds; this extra mixing has been suggested by
many authors (see below).  Several abundance peculiarities besides the
oxygen problem are explained by introducing CBP during at least the
red giant branch (RGB) phase of evolution.  Specifically, observed
patterns of $^7$Li production and destruction, and of $^{13}$C
production on the RGB require CBP (Gilroy \& Brown 1991; Charbonnel
1995; Boothroyd \& Sackmann 1999; Gratton et al. 2000).  CBP also
explains the evolution in time of the carbon isotope abundances along
the RGB (Denissenkov \& Weiss, 1996; Cavallo, Sweigart \& Bell 1998).
The actual mechanism that causes this mixing is not known.  It is
often ascribed to the shear instability in differentially rotating
stars (Zahn 1992; Maeder \& Zahn 1998), but other mechanisms, such as
meridional circulation and convective overshoot (Sweigart \& Mengel
1979; Herwig et al. 1997 and references therein), are available and equally
likely.  The specific mechanism will determine the actual depth and
rate of mixing, but should not otherwise matter.

Deep mixing processes were shown to be especially important in
low-metallicity stars, where deep mixing may involve relatively hot
H-burning cycles.  CBP was therefore suggested (e.g.  by Langer,
Hoffman \& Sneden 1993, Weiss, Denissenkov \& Charbonnel 2000) as an
explanation for a collection of CNO, Ne-Na and Mg-Al anomalies
observed in low-metallicity stars (reviewed in e.g. Kraft 1994, Da
Costa 1998).  These studies showed that when CBP becomes deep enough
to involve the H shell itself, the stellar evolution is significantly
altered by extensive He enrichment of the envelope (Sweigart 1997).
Denissenkov \& Weiss (2001), investigating elemental abundance
anomalies in the globular clusters $\omega$ Cen and M4, found that at
least some of the observed variations in [Al/Fe] could result from CBP
occurring in RGB stars.  These workers used a diffusive model of extra
mixing with rate and depth of mixing determined to fit observational
trends. They demonstrated the possibility of $^{26}$Al production in
RGB stars undergoing CBP, at least in low-metallicity stars and with
particular parameters.  Very recently, observations of planetary
nebulae were shown to require further addition of CN-cycle products
into the envelope during the third dredge-up phase, even in stars with
masses that are too low to permit HBB (P\'equignot et al. 2000).  The
destruction of $^3$He that should accompany CBP would also help
explain the apparently weak evolution of the Galactic $^3$He abundance
since the big bang (Wasserburg, Boothroyd \& Sackmann 1995; Balser et
al. 1998; Gloeckler \& Geiss 1998; Bania, Rood \& Balser 2002).

For all the above reasons, CBP is now generally accepted as a
significant process occurring in a second mixing episode late in the
evolution of red giants (see e.g. Boothroyd \& Sackmann 1999; Gratton
et al. 2000).  The main restriction on CBP is that there be no abrupt
increase in mean molecular weight anywhere between the convective
envelope and the hydrogen shell to prevent mixing (Charbonnel, Brown
\& Wallerstein 1998; Boothroyd \& Sackmann 1999). This condition is
known to be satisfied in $M<2\ms$ stars on the RGB, at luminosities
above that of the ``bump'' in the RGB luminosity function.  It should
also be satisfied throughout the subsequent AGB phase -- and certainly
during the thermally-pulsing AGB (TP-AGB) phases -- for all low-mass
stars.  From all these considerations CBP therefore appears to be both
possible and necessary in the evolution of low-mass stars.

\section{The Model}
\label{sec:stationary}

\begin{figure}
\epsscale{0.6}
\plotone{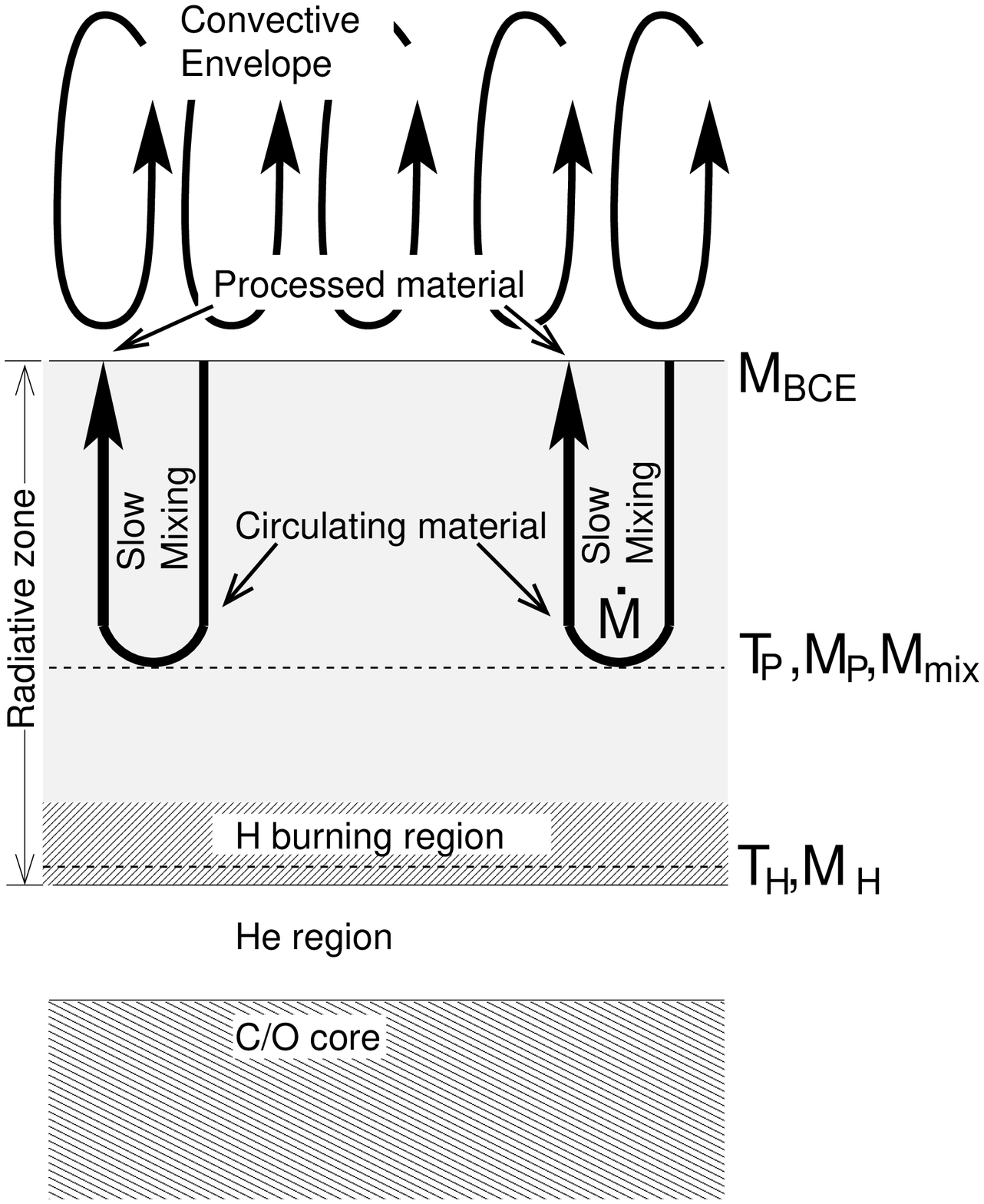}
\caption{Schematic diagram of the cool bottom processing model.
Material taken from the envelope circulates slowly down into the
radiative zone, where it undergoes nuclear processing at temperatures
near $T_P$.  It then returns to the envelope, where it is rapidly
mixed with the other envelope material.  Relative thicknesses of the
various layers do not reflect physical dimensions.}
\label{fig:cartoon}
\end{figure}

The general structure and evolution of AGB stars have been studied
extensively (for a thorough introduction see Lattanzio
2001\footnote{John Lattanzio's online stellar evolution tutorial
available at http://www.maths.monash.edu.au/~johnl/stellarevoln/.}).
The structure of a low-mass star during this phase of evolution is
schematically as follows (see Fig.  \ref{fig:cartoon}): in the center
is a degenerate carbon/oxygen core.  Outside this is a helium region
which cannot support steady nuclear burning.  The next layer is a thin
($< 10^{-4}\ms$) region where hydrogen burning takes place, providing
essentially all of the star's luminosity.  The outward advance of the
H shell adds to the He region, which periodically undergoes nuclear
burning (``thermal pulses'') in short ($\sim 200$ yrs) bursts
separated by long ($\sim 30,000$ yrs) intervals.  The H shell and the
region immediately above it ($< 10^{-3}\ms$ in total) are stable
against convection, and we call these together the ``radiative zone.''
Finally, the outer region of the star is a large, fully convective
envelope.  When mass loss is included as described below, this
envelope has mass $\sim 0.9\ms$ at the start of the TP-AGB and shrinks
as the H shell advances outward and as the star sheds envelope
material in a wind.  Late in the AGB evolution, each thermal pulse is
followed by a brief disruption of the structure just described, during
which material that has seen nuclear processing in the H and He shells
is mixed into the envelope.  This mixing is called ``third dredge-up''
(see review by BGW99).  In the above description, the radiative zone
contains no circulation.  We will treat the problem of extra mixing
following Wasserburg, Boothroyd \& Sackmann (1995; hereafter WBS).
Here a slow circulation is added by imposing transport of envelope
material at a mass flow rate \mdot\ down to some depth inside the
radiative zone specified by temperature $T_P$ and then back to the
envelope.  Typically, a small amount of nuclear processing takes place
in the circulating material.

\begin{figure}
\epsscale{1}
\plotone{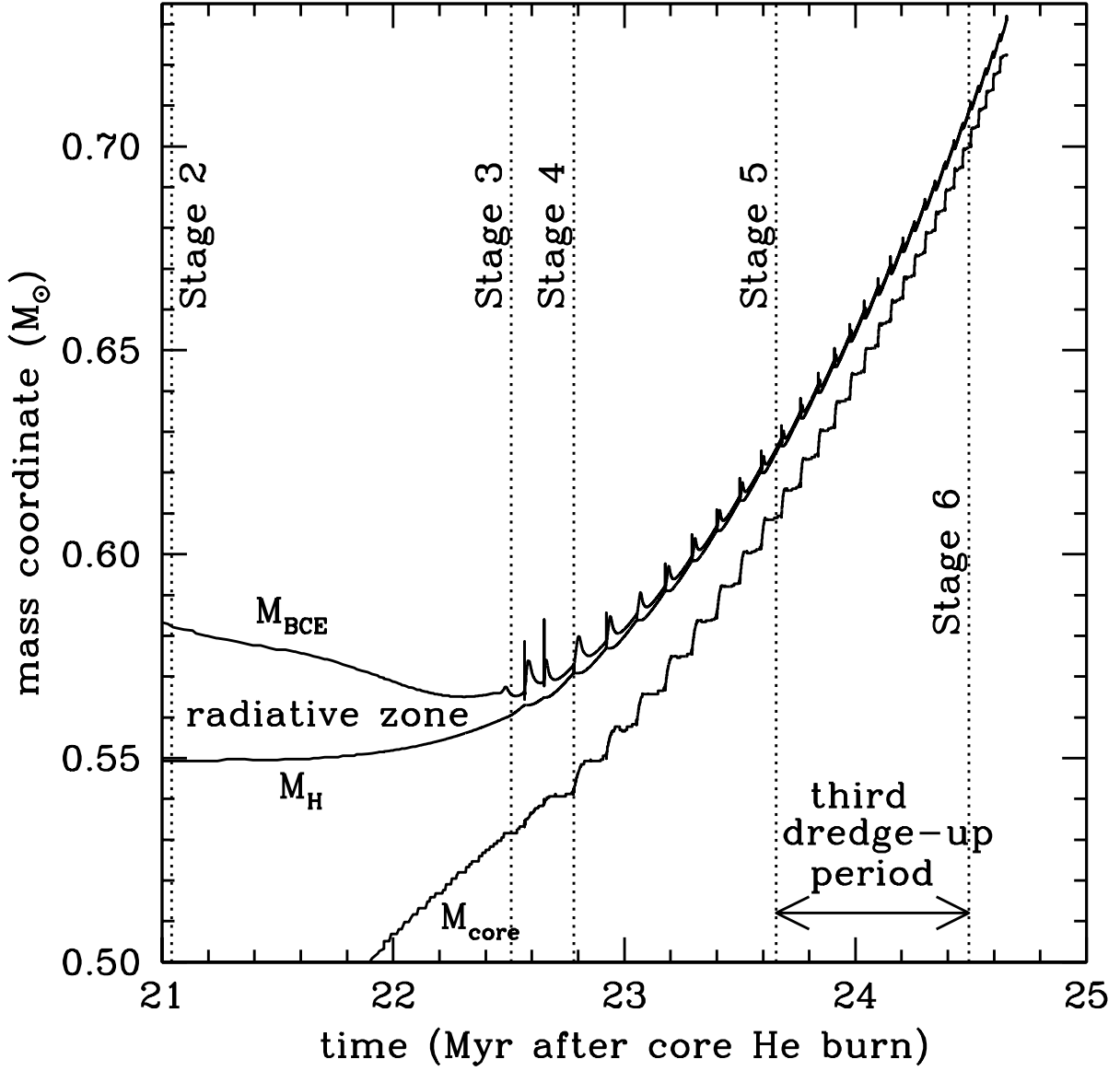}
\epsscale{0.6}
\caption{Evolution of the boundary of the C/O core $M_\mathrm{core}$,
location of the H shell $M_H$, and convective-envelope boundary
$M_{BCE}$ for the 1.5\ms\ solar-metallicity model of Straniero et al.
(1997) [SEM(1.5)], as functions of time since the He flash.
Vertical lines indicate times selected to provide radiative-zone
structures for our CBP calculations (cf. Table
\protect\ref{tab:stages}).}
\label{fig:tpagb}
\end{figure}

We use a numerical model of an AGB star with solar initial composition
and 1.5\ms\ initial mass.  The model was computed by Straniero et
al. (1997) using the FRANEC code to follow the full evolution of the
star from the zero-age main sequence through the TP-AGB phase, with no
mass loss.  The corresponding loss of envelope mass was subsequently
computed using the Reimers (1975) prescription as discussed by Busso
et al. (2001).  The choice $\eta=1$ for the free parameter of the
Reimers formula was adopted.  Details of the mass loss and dredge-up
are relatively unimportant for our conclusions: runs of the code at a
fixed envelope mass of 0.5\ms\ produced the same results to $\sim
10\%$, and variations in dredge-up efficiency will be discussed below.
The detailed evolution on the AGB of the above model is summarized in
Fig. \ref{fig:tpagb} and Table \ref{tab:stages}.  We will refer to the
detailed structure and evolution of this model, including both
dredge-up and mass loss, as the stellar evolution model, SEM(1.5).
For our calculations, we assume that extra mixing may be modelled by
studying circulation inside the radiative-zone structures previously
computed in the SEM(1.5) model.  The nuclear processing in CBP depends
mainly on the variation of temperature $T$ and density $\rho$ as
functions of position within these radiative-zone structures, and of
time $t$.  These structures are the environment within which the
circulating material moves and is processed.  The SEM(1.5) model also
provides us with the timescales for AGB evolution, the rate of mass
loss from the envelope, and the changes to envelope composition due to
third dredge-up (TDU).  The SEM(1.5) model exhibits TDU after the 10th
thermal pulse, and after each subsequent thermal pulse until the 26th
(the interval from Stage 5 to Stage 6 in Fig. \ref{fig:tpagb} and
Table \ref{tab:stages}; two subsequent pulses produce negligible
dredge-up).  This dredge-up brings large quantities of newly-produced
carbon and of s-process nuclides to the envelope, and as a result, the
envelope of SEM(1.5) with no CBP has C/O$>1$ from the 22nd pulse
onward.  The envelope begins CBP with a composition reflecting earlier
first dredge-up, as discussed below for individual nuclides.

It is useful in discussing CBP to distinguish between three types of
material.  ``Envelope material'' (E) makes up the convective envelope
and has abundance (in moles per gram) $Y_i^E$ of nuclide
$i$. ``Circulating material'' (C) is that material moving through the
radiative zone.  It starts out as a representative sample of the
envelope at a given time, and then moves down through the radiative
zone and back as shown in Fig. \ref{fig:cartoon}. Abundances $Y_i^C$
in the circulating material are changed by $(p,\gamma)$ and
$(p,\alpha)$ reactions, as well as beta decays along the flow path.
The circulating material just before returning to the envelope is the
``processed material'' (P), with composition denoted $Y_i^P$.  We
assume that mixing of processed material with the convective envelope
is instantaneous.  

The numerical CBP model follows the envelope abundances as material
descends into the radiative zone and is replaced by newly-processed
material.  This corresponds to integrating
\begin{equation}
\dot{Y}^E_i = \frac{\mdot}{M_E}\left(Y^P_i-Y^E_i\right)
\end{equation}
as $M_E$, $Y^P_i$, and $Y^E_i$ evolve together.  The quantity $Y^P_i$
is computed by following the position of a parcel of material as it
starts at the bottom of the envelope, circulates downward through the
SEM(1.5) stellar structure, and finally returns to the envelope
with composition $Y^P_i$.  The stellar structure and \mdot\ specify as
functions of time the conditions of $T$ and $\rho$ seen by this
material, and the numerical code integrates a nuclear reaction network
(rates from Angulo et al. 1999; discussed in Appendix \ref{sec:rates})
with these changing conditions.  We do not calculate the detailed
evolution by following every detail of development of the radiative
zone with time.  For a given calculation, we choose one of the times
from the SEM(1.5) evolution listed in Table \ref{tab:stages} (we
call this a ``stage''), and we hold the radiative-zone structure
constant as specified by this stage.  This amounts to an approximation
that the processing is independent of the evolution of the radiative
zone for the duration of CBP.  A check of this assumption is given in
Sec. \ref{sec:consistency}.  We consider the dominant extra processing
to take place when the H shell temperature $T_H$ (defined as the
temperature where the maximum energy generation takes place) exceeds
$\log T_H\gtrsim 7.7$ and when TDU is active.  This is restricted to
the last $2\times 10^6$ yrs of AGB evolution of the 1.5\ms\ star.
When considering CBP during the interval between Stages 5 and 6, we
use the radiative-zone structure from Stage 5, and we include TDU by
interrupting the evolution at the times of the thermal pulses and
altering the envelope abundances to reflect dredge-up as computed for
SEM(1.5).  The CBP code is re-started after each pulse with the new
composition as the intial state.  We refer to the results of this
calculation as the ``numerical model.''  We will also present
analytical arguments to interpret the output.  After an initial
examination of the conditions necessary for $^{26}$Al production and
incorporation into grains, we confine ourselves to evolution from
Stage 5 to Stage 6, when TDU is active, and we use the radiative-zone
structure from Stage 5.  All of the results from the numerical model
shown in the figures correspond to these calculations.

A fully self-consistent model of CBP using this ``conveyor belt''
model would involve calculating the abundances of the nuclides in the
envelope at all times and following the evolution of stellar structure
as conditions in the radiative zone change, as the H-shell advances
into regions that have been processed by CBP, and as dredge-up and
mass loss change the envelope.  The most extreme assumption that we
have made is that a radiative-zone structure from a fixed time
provides a reasonable approximation to the radiative-zone structure
for the next few $10^5$ yrs.  We find below that this is true during
the TP-AGB (excluding very brief interruptions by thermal pulses), and
that C, N and O isotopes are insensitive to the specific
radiative-zone structure chosen.  We will show that for $^{26}$Al
there is a strong dependence on this structure, dominantly due to the
evolving $T_H$.  It will be shown that there are nonetheless clear
general rules for $^{26}$Al that adequately describe the system.  Note
that the conditions of $T$ and $\rho$ in the envelope play no role in
this problem as no nuclear processing occurs there.

The radiative zone is only $8.6\times 10^{-4}\ms$ thick at Stage 5, so
if the envelope has a convective overturn time of $\sim 1$ yr, and the
radiative zone turns over much less frequently than the envelope, then
$\mdot < 10^{-4}\ms$/yr, roughly.  The other important constraint is
that stellar energy generation not be disrupted too much, or
approximately that $T_P< T_H$.
In principle, \mdot\ does not completely specify the speed of matter
circulation.  The speeds of the upward and downward streams may be
expressed as in WBS by $f_u$ and $f_d$, respectively the fraction of
the area at fixed radius occupied by the upward and downward
streams. We have fixed both parameters at $f_u=f_d=f=0.5$, and they
will not have any significant effect for fixed $\fup+\fdown$ (see
WBS). For each calculation we fix a value of $\mdot$ and a maximum
depth of penetration in the radiative zone corresponding to a
temperature $T_P$.  There are no other independent parameters in the
calculation.

\subsection{$^{26}$Al}
\label{sec:al26}

We first consider $^{26}$Al which is the simplest case.  The
long-lived ground state of $^{26}$Al is produced in the circulating
material by the process $^{25}{\rm Mg}(p,\gamma)^{26}{\rm Al}$ and
destroyed by $^{26}{\rm Al}(p,\gamma)^{27}{\rm Si}$.  The $^{26}$Al
abundance $Y_{26}^C$ in a sample of circulating material starts with
the current composition of the envelope and then evolves during the
time that it circulates through the radiative zone according to
\begin{equation}
\label{eqn:al26}
\dot{Y}^C_{26} = Y^C_{25}\lambda[^{25}{\rm Mg}(p,\gamma)]
                 -Y^C_{26}\lambda[^{26}{\rm Al}(p,\gamma)],
\end{equation}
where $\lambda[{\rm rxn}]=\rho N_A\langle\sigma v\rangle Y_H^C$ for
the reaction labeled ``rxn'', $\rho$ is mass density, $N_A$ Avogadro's
number, $\langle\sigma v\rangle$ the thermally-averaged reaction rate,
$Y_{25}^C$ the $^{25}$Mg abundance, and $Y_H^C$ the hydrogen
abundance.  The quantities on the right-hand side of this expression
evolve as the material's position in the radiative zone changes.  The
destruction term is negligible for two reasons: 1) for the
presently-recommended rates (Appendix \ref{sec:rates} and Angulo et
al.  1999), the destruction cross section is two orders of magnitude
smaller than the production cross section, and 2) $Y^C_{26}$ never
builds up to a level comparable to $Y^C_{25}$.  The rate of $^{26}$Al
decay during the time spent circulating through the radiative zone is
negligible.

For most values of \mdot\ and for $T_P < T_H$, the $^{26}$Al
production rate is low enough that only a small fraction of the
$^{25}$Mg burns on a single pass through the radiative zone.  As a
result, $Y_{25}^C$ is almost constant, and the amount of $^{26}$Al in
the processed material is approximately
\begin{eqnarray}
\label{eqn:al26approx} Y^P_{26}(t_f) &=& Y^E_{26}(t_i)
+\int_{t_i}^{t_f} Y^C_{25}\lambda[^{25}{\rm Mg}(p,\gamma)]\, dt
\\
    Y^P_{26}(t_f) &=& Y^E_{26}(t_i)+2 f
\mdot^{-1}Y^E_{25}Y_H^E\int_{M_P}^{M_{BCE}} \rho N_A\langle\sigma
v\rangle\, dm,
\label{eqn:mdotscaling}
\end{eqnarray}
where we have taken advantage of the relationship between \mdot\ and
the time-dependent position of the circulating matter to write the
integral in mass coordinates with no explicit reference to time. $M_P$
is where $T=T_P$ and $M_{BCE}$ is the bottom of the convective
envelope (see Fig. \ref{fig:cartoon}) -- the time integral begins at
the time $t_i$ when the circulating material leaves the envelope and
ends at time $t_f$ when it returns to the envelope. (The factor of 2
arises because the circulating material passes each point twice: once
on the way down, and once on the way back up.) Note that
$Y^P_{26}(t_f)- Y^E_{26}(t_i)$ does not change with time, except to
the extent that stellar evolution changes the path integral.

The abundance of $^{26}$Al in the envelope reflects the addition
of processed material with the $^{26}$Al abundance $Y_{26}^P$ to
the envelope of mass $M_E$, competing against free decay with mean
lifetime $\tau_{26}=1.0\times 10^6$ yr and removal of material
into the downward stream:
\begin{eqnarray}
\dot{Y}_{26}^E & = & \frac{\mdot}{M_E} \left(Y_{26}^P - Y_{26}^E\right)
                      - \tau_{26}^{-1} Y_{26}^E\\
                & = & 2 f M_E^{-1}Y^E_{25}(0) Y_H^E \int_{M_P}^{M_{BCE}} 
                     \rho N_A\langle\sigma v\rangle\, dm
                      - \tau_{26}^{-1} Y_{26}^E
\label{eqn:prodrate-1}\\
                & = & P - \tau_{26}^{-1} Y_{26}^E,
\label{eqn:prodrate}
\end{eqnarray}
where the last two lines define the production rate $P$ per unit
mass of the envelope. Note that all factors of \mdot\ have
cancelled: the amount of processed material is proportional to
\mdot, but the time it spends at high temperature is proportional
to $\mdot^{-1}$.  If the envelope mass decreases with time and the
reaction path integral over the circulation stays constant, then
$P$ will increase with $1/M_E(t)$.

\begin{figure}
\epsscale{1}
\plottwo{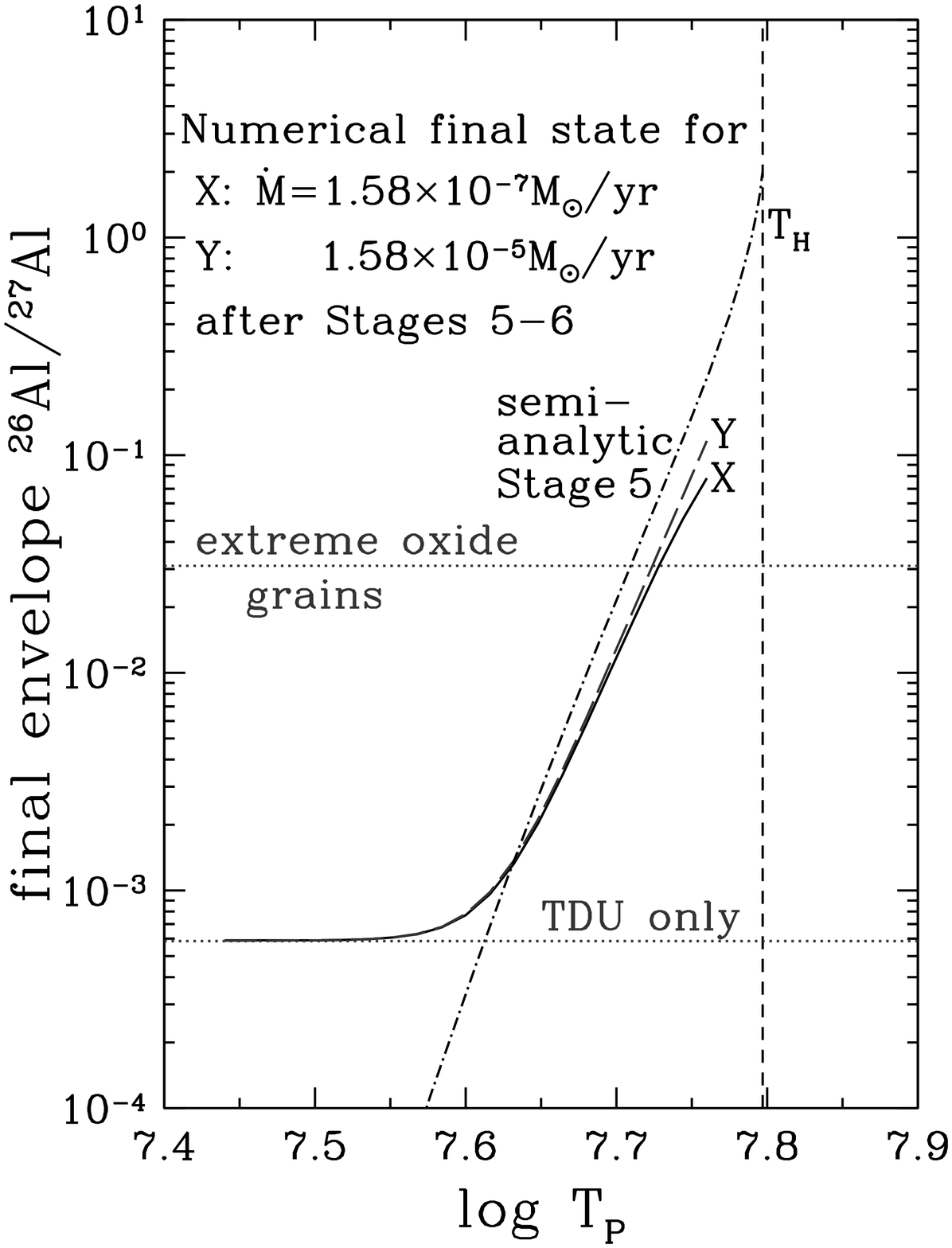}{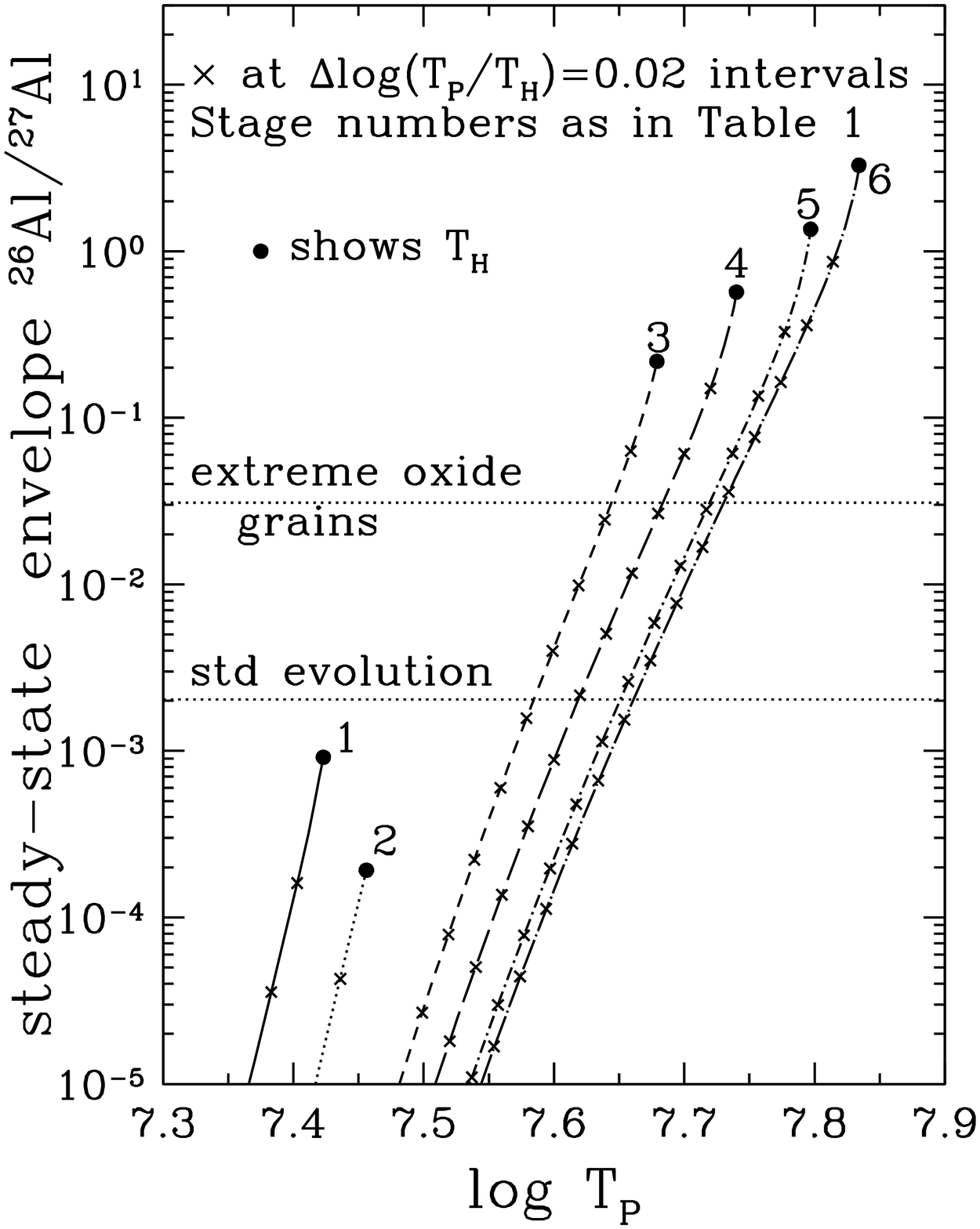}
\epsscale{.7}
\caption{a) Final $^{26}$Al/$^{27}$Al as functions of $T_P$ from the
numerical model running for the $8.5\times 10^5$ yr period during TDU
(Stage 5 to Stage 6), compared with the corresponding semi-analytic
result for the CBP steady-state abundance (curve 5 of panel b).
``Extreme oxide grains'' indicates the highest value found for a
meteoritic oxide grain.  b) Envelope $^{26}$Al/$^{27}$Al at steady
state ( Eqn. \protect\ref{eqn:prodrate}).  The production rate $P$
changes as the star evolves; each curve shows the value of
$\tau_{26}P/Y^E_{27}$ for a particular choice of radiative zone
structure taken from the stellar evolution model.  Curves are labelled
according to stage numbers in Table \protect\ref{tab:stages}.  The end
of each curve indicates $T_H$ at that time.  The ``std evolution''
line indicates the level found for TDU production of $^{26}$Al by
BGW99.}
\label{fig:integral}
\end{figure}

Calculations were done in the numerical model using the radiative-zone
structure of Stage 5, with CBP occurring roughly for the duration of
third dredge-up (from Stage 5 until Stage 6).  The results for
$^{26}$Al are shown in Fig. \ref{fig:integral}a as a function of
$T_P$.  During this interval, $^{26}$Al/$^{27}$Al $= 6\times 10^{-4}$
is added gradually by normal third dredge-up, and this amount is shown
as the ``TDU only'' line.  As argued above
(cf. Eqn. \ref{eqn:prodrate-1}), there is no significant dependence on
$\mdot$.  We see that for $\log T_P \lesssim 7.6$ (0.2 dex below
$T_H$) there is no significant $^{26}$Al production. The term $\log
\left(^{26}{\rm Al}/^{27}{\rm Al}\right)$ rises almost linearly with
$\log T_P$ as $T_P$ increases toward $T_H$. It is evident that CBP
penetrating to temperatures above $T_P=10^{7.6}$K during this phase of
evolution can readily produce $^{26}$Al/$^{27}$Al as high as found in
extreme examples of circumstellar oxide grains.

In order to understand these numerical results, we consider analytic
solutions to the approximate expression in
Eqn. \ref{eqn:prodrate-1}. Table \ref{tab:stages} and
Fig. \ref{fig:tpagb} show the selected stages of AGB evolution on
which we will now focus.  Time zero is taken as the state when He core
burning is just finished, and qualitative descriptions of the
evolutionary stages are given in Table \ref{tab:stages} along with
characteristic parameters of the radiative zone.  For the
semi-analytic treatment, we neglect mass loss and dredge-up, and we
assume that the stellar structure (i.e., the structure at a selected
stage from Table \ref{tab:stages}) does not change with time.  Using
the solution to Eqn. \ref{eqn:prodrate} for a constant-mass envelope
and a selected time-independent radiative-zone structure, we obtain
\begin{equation}
Y_{26}^E=Y_{26}^E(0) e^{-t/\tau_{26}}
           + \left[1-e^{-t/\tau_{26}}\right] \tau_{26}P.
\label{eqn:solution}
\end{equation}

Let us now consider the steady-state abundance of $^{26}$Al that could
be achieved in an AGB star fixed at the structure from a particular
evolutionary stage.  The results for Stage 5 are shown in
Fig. \ref{fig:integral}a and are in good agreement with the
corresponding numerical calculation.  We consider the following
additional times in the evolution: 1) The stellar structure at the
immediate end of core He burning; 2) the middle of the early AGB; 3)
just before the first thermal pulse; 4) just after the second mild
pulse, etc.; and finally 6) just before the twenty-sixth thermal pulse
(see Table \ref{tab:stages} and Fig. \ref{fig:tpagb}). The
$^{26}$Al/$^{27}$Al ratios at steady state ($\tau_{26}P/Y^E_{27}$) for
each of these stellar stages are shown in Fig \ref{fig:integral}b.
There is no dependence on \mdot\ because of the cancellation noted
above.  The end points of the curves correspond to the extreme
limiting case of $T_P = T_H$.  By the onset of the first thermal
pulse, the production rates rise rapidly and $^{26}$Al production is
possible only at these late stages.  The critical parameter is $T_P$,
which is not a priori known.  Values of $\log T_P\gtrsim 7.6$ are
required to produce substantial amounts of $^{26}$Al.  Inspection of
Fig. \ref{fig:integral}b shows that to achieve the same value of
$^{26}$Al/$^{27}$Al requires different in $T_P$ for each choice of
stellar structure (e.g., $^{26}$Al/$^{27}$Al$=0.03$ requires $\log
T_P= 7.64$ at Stage 3, 7.68 at Stage 4, 7.72 at Stage 5, and 7.73 at
Stage 6).  However, if we had chosen $T_P/T_H\sim$ constant for each
of the last three stages rather than fixing $T_P$, then the
$^{26}$Al/$^{27}$Al production rate would be approximately constant.
(See section \ref{sec:consistency} below.)

\begin{table}
\caption{Stages of AGB evolution considered for CBP.  ``Start time''
is in years after the end of core He burning.  $\Delta M$ is the
thickness of the radiative zone starting at the bottom of the H shell.
The numerical model is based on Stage 5, assuming a constant structure
for the radiative zone for the interval until Stage 6.}
\label{tab:stages}
\begin{tabular}{lclclc}
Stage & Start Time (Myr) & Description  & $\log T_H$ & $\Delta M$ (\ms) &
$M_E\ (M_\odot)$\\
\hline
1    &         0    & end core He burn     &  7.423  &   0.376   &  0.587 \\
2    &   $21.04$    & before H-shell re-ignition &  7.456  &   0.0329  &  0.900 \\
3    &   $22.47$    & before 1st pulse     &  7.679  & $5.18\times 10^{-3}$
& 0.862\\
4    &   $22.78$    & after 2 mild pulses  &  7.740  & $2.02\times 10^{-3}$
& 0.887\\
5    &   $23.66$    & before 10th pulse (start TDU)    & 7.797
& $8.65\times 10^{-4}$   & 0.741 \\
6    &   $24.49$    & after 26th pulse (end TDU) & 7.834
& $3.17\times 10^{-4}$   & 0.242 \\
\end{tabular}
\end{table}

In general, high levels of $^{26}$Al/$^{27}$Al can be achieved by CBP,
essentially independent of $\mdot$, but completely dependent on $T_P$.
Thus, $^{26}$Al/$^{27}$Al is a measure of the depth of penetration,
independent of \mdot.  Significant $^{26}$Al production can occur over
all the late stages, but we expect it to be dominated by production
over the last $\tau_{26}$ of time before the envelope is lost.  It can
be seen in Fig. \ref{fig:integral}a that the numerical model is very
close to the simple analytical calculation.  Most of the small
discrepancy arises because the TDU period lasts only for $8.5\times
10^5$ yr, not long enough to reach steady state.  Overall, the
agreement is astoundingly good.

In summary, significant $^{26}$Al/$^{27}$Al can result from CBP late
on the AGB.  The exact amount of $^{26}$Al produced depends on the
depth of circulation as expressed by $T_P$ and the duration of CBP
(limited by the AGB lifetime), but only very weakly on the circulation
rate \mdot.  To produce $^{26}$Al/$^{27}\mathrm{Al}\gtrsim 4 \times
10^{-3}$ just before the star ejects its envelope, it appears
necessary that $\log T_P> 7.67$.  There is only small $^{26}$Al
production in the early AGB phase and we infer that this also applies
to any CBP during the preceding RGB phase because of the low $T_H$ at
that time.

\subsection{Oxygen}
\label{sec:oxygen}

We now examine the implications for the oxygen isotopes of the
conditions that produce $^{26}$Al. In contrast to $^{26}$Al, $^{18}$O
is not produced but is destroyed.  First dredge-up during the RGB
phase provides the envelope with $^{18}$O/$^{16}$O that is decreased
from its initial value due to destruction during the main-sequence
phase.  This has been shown by Dearborn (1992), who recognized the
significance of $^{18}$O and $^{17}$O as indices of interior stellar
temperatures and mixing processes. These results have been confirmed
by El Eid (1994), Boothroyd \& Sackmann (1999), Gallino et al. (1998)
and others. In contrast, the isotope $^{17}$O is produced in the
interior and is then enriched in the envelope above the initial value
as a result of first dredge-up.  Neither $^{18}$O nor $^{17}$O is
affected by thermal pulses. The major isotope in the envelope,
$^{16}$O, is essentially unchanged throughout stellar evolution.

\begin{figure}
\plotone{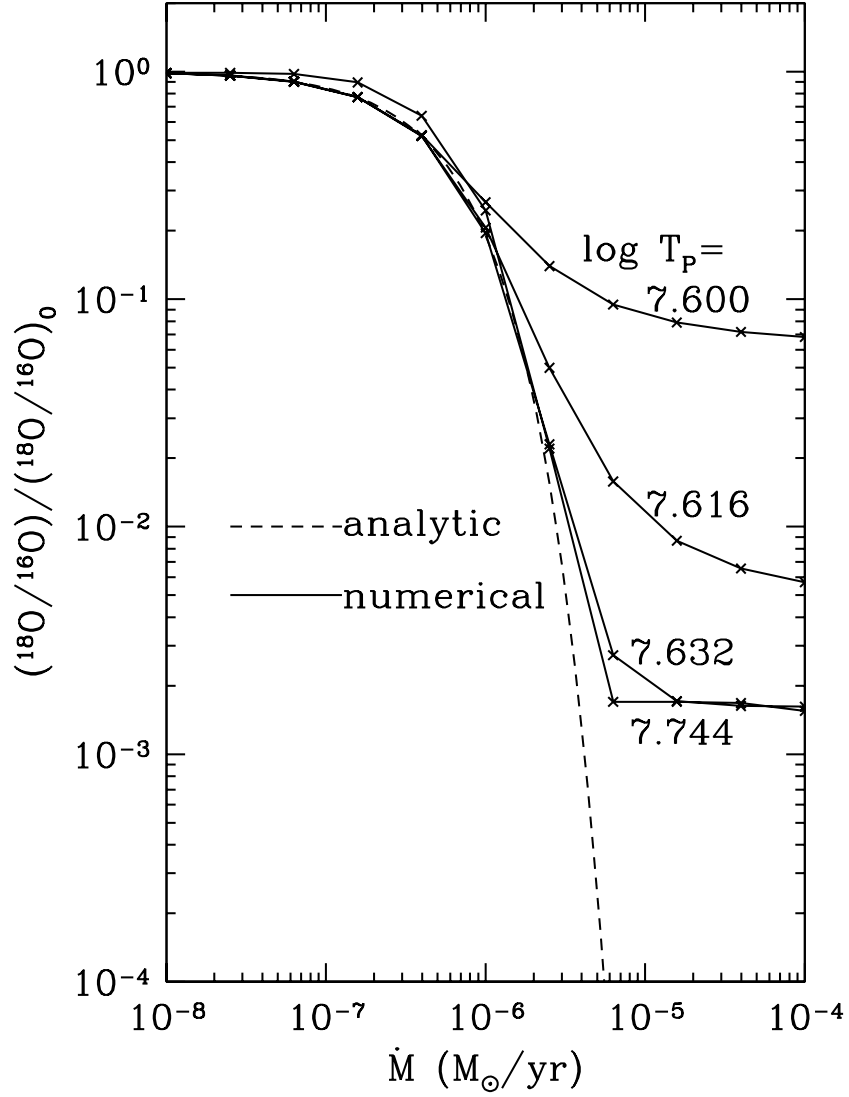}
\caption{Fraction of $^{18}$O surviving in the final stellar envelope,
as a function of \mdot, after $8.5\times 10^{5}$ yr.  The ``analytic''
curve is based on a stellar envelope of 0.5\ms\ and complete
destruction of $^{18}$O in the circulating material
(Eqn. \ref{eqn:decay}) .  The solid curves correspond to processing
for the same amount of time in the numerical model, at four different
values of $T_P$.}
\label{fig:mdotvs18}
\end{figure}

\begin{figure}
\plotone{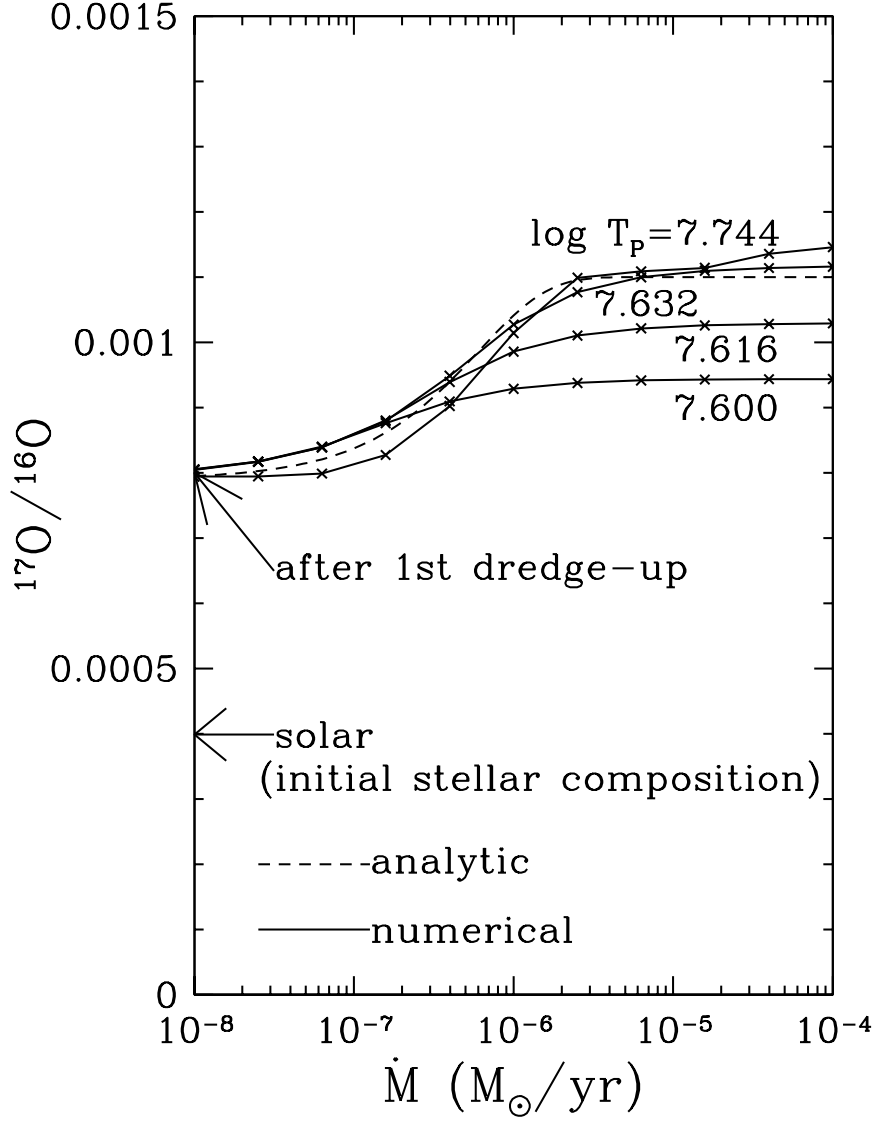}
\caption{$^{17}$O/$^{16}$O ratio in the final stellar envelope after
processing throughout the TDU period.  The analytic curve
(Eqn. \ref{eqn:mixing}) reflects an initial envelope with composition
set by first dredge-up, mixing with processed material with the
composition $^{17}$O/$^{16}$O$=0.0011$.  The other curves are results
from the corresponding numerical model, at 4 different values of
$T_P$.}
\label{fig:mdotvs17}
\end{figure}

The results of the numerical model for $^{18}$O/$^{16}$O and
$^{17}$O/$^{16}$O are shown in Figs. \ref{fig:mdotvs18} and
\ref{fig:mdotvs17} as functions of $\mdot$.  The curves show the final
envelope composition after using the fixed radiative-zone structure of
Stage 5 and processing through the length of the TDU period.  Curves
for various values of $\log T_P$ are given.  It can be seen that for
$\mdot< 10^{-7} \ms$/yr, there is no substantial change in $^{18}$O
for all $T_P$.  This is because very little material is processed over
the lifetime of the AGB for these mixing rates.  For $\mdot \gtrsim
3\times 10^{-7} \ms$/yr there is a sharp drop-off with a magnitude
depending on $T_P$.  For rates above $3\times10^{-6}\ms$/yr, the
destruction is almost complete when $\log T_P > 7.62$.  The horizontal
parts of the curves at higher $T_P$ are due to a small amount of
$^{18}$O production balancing the destruction.  For $^{17}$O there is
a rise to higher $^{17}$O/$^{16}$O by a factor of $\sim 1.5$
(dependent on $T_P$) above the value after first dredge-up.  We find
that if $\log T_P$ exceeds a low threshold, the calculation of the
final composition as a function of \mdot\ does not change
significantly.  This threshold is 7.62 at Stage 5 and 7.60 at Stage 3.
Therefore, the results are robust and a calculation of the full
evolution with continuously changing structure and CBP will not
significantly alter the results found here by assuming a fixed
radiative zone.

In order to understand the results, we examine the transport equations
as was done for $^{26}$Al.  The basic results can be obtained almost
quantitatively from an analytic treatment. As long as $\log T_P >
7.62$, the evolution of $^{18}$O in CBP is characterized by rapid
destruction.  (This applies for Stage 5; other stages are
characterized by analogous thresholds, never lower than 7.60 or higher
than 7.65.)  This destruction occurs via $^{18}{\rm
O}(p,\alpha)^{15}{\rm N}$, with $^{18}$O abundance in the circulating
material evolving according to
\begin{equation}
\dot{Y}_{18}^C = - \lambda[^{18}{\rm O}(p,\alpha)] Y_{18}^C.
\end{equation}
(Production via $^{17}{\rm
O}(p,\gamma)^{18}\mathrm{F}(\beta^+\nu)^{18}{\rm O}$ is negligible).
The $^{18}$O abundance in the processed material returning to the
envelope is the time integral of this equation along the entire path
taken by the material through the radiative zone, given by
\begin{eqnarray}
Y_{18}^P & = & Y_{18}^E \exp\left\{-\int_{t_i}^{t_f} 
                \lambda[^{18}{\rm O}(p,\alpha)]\,dt\right\}\\
          & = & Y_{18}^E \exp\left\{-2f\mdot^{-1}Y_H^C\int_{M_P}^{M_{BCE}}
                \rho N_A \langle\sigma v\rangle\,dm\right\}.
\end{eqnarray}
The exponential factor is typically very small for even modest values
of $T_P$; for example, this expression gives $Y_{18}^P=0.007\,
Y_{18}^E$ if $\log T_P=7.67$ and $\mdot=10^{-6}$ \ms/yr using the
radiative zone of Stage 5.  The result is $^{18}$O/$^{16}$O as low as
$10^{-5}$ in the processed material. Higher $T_P$ or lower \mdot\
would further reduce $Y_{18}^P$, as would processing slightly earlier
in the TP-AGB evolution when the threshold for destruction is lower
(analogously with $^{26}$Al production). For example, using the
radiative zone of Stage 5 with $\log T_P=7.69$ and $\mdot=10^{-6}$
\ms/yr gives $Y_{18}^P=4.2\times 10^{-5} Y_{18}^E$.  Using the
radiative zone at Stage 3 with $\mdot=10^{-6}$ \ms/yr, only $\log
T_P=7.61$ is needed to produce $Y_{18}^P=0.007\, Y_{18}^E$.

Because $^{18}$O destruction in the processed material is so thorough,
the envelope abundance of $^{18}$O reflects only the removal of
envelope material and its replacement with material containing
essentially no $^{18}$O:
\begin{eqnarray}
\dot{Y}_{18}^E  
& \simeq & -\frac{\mdot}{M_E} Y^E_{18}.
\end{eqnarray}
The result is exponential decay of $^{18}$O/$^{16}$O with time, so if
$M_E$ is constant,
\begin{equation}
\frac{Y_{18}^E}{Y_{16}^E} = \frac{Y_{18}^E(0)}{Y_{16}^E}\exp(-
\mdot t / M_E). 
\label{eqn:decay}
\end{equation}

In this analytical treatment, the destruction of $^{18}$O in the
processed material is taken to be almost complete.  This will apply to
all stages of the evolution above the threshold $T_P$ discussed above.
Hence the resulting $^{18}$O/$^{16}$O in the envelope at the end of
the AGB is given by Eqn. \ref{eqn:decay} where the only parameters are
$\mdot, M_E$, and the time scale for AGB evolution. This is shown for
$t=8.5\times 10^5$ yr (the length of the TDU period) and $M_E=0.5\ms$
as the dashed curve in Figure \ref{fig:mdotvs18}. It can be seen that
the semi-analytic treatment is almost indistinguishable from the
numerical model over most of the range of interest. The flat region at
$\mdot > 10^{-5}\ms$/yr would have been obtained if we had not
neglected the small $^{18}$O production.

\begin{figure}
\plotone{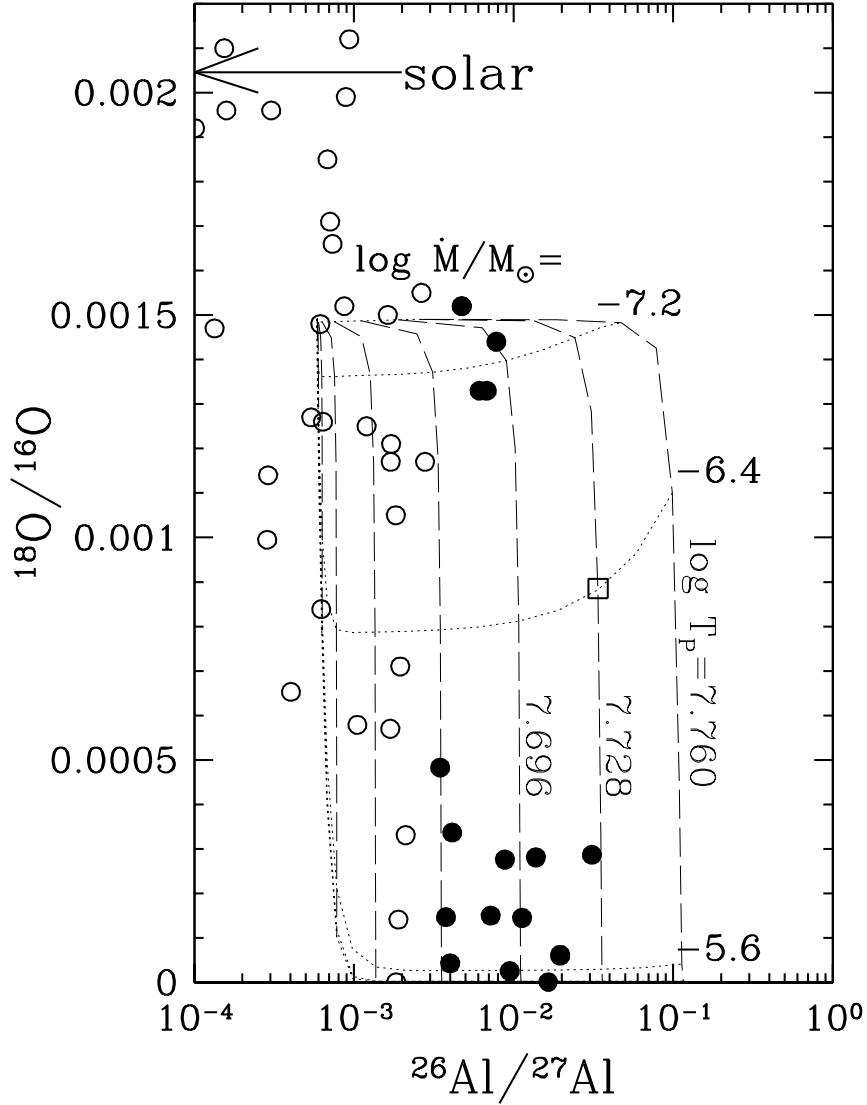}
\caption{Final envelope compositions from the numerical model with the
Stage 5 radiative-zone structure.  The grid consists of curves of
constant \mdot\ (dotted lines) and constant $T_P$ (dashed lines).
Superimposed over the grid are the oxide grain data summarized in Choi
et al. (1998); filled circles are those with
$^{26}$Al/$^{27}$Al$>2\times 10^{-3}$.  In principle, one can read
values of \mdot\ and $T_P$ off of this and subsequent figures; for
example, a grain with $^{26}$Al/$^{27}$Al$=3.5\times 10^{-2}$ and
$^{18}$O/$^{16}$O$=0.00087$ (at the square) corresponds to the final
product of CBP with $\log T_P=7.728$ and $\log (\mdot/\ms)=-6.4$.}
\label{fig:o18endpoints}
\end{figure}

Because of the low temperature at which $^{18}$O is destroyed, it is
evident that low $^{18}$O/$^{16}$O in the envelope is not necessarily
related to $^{26}$Al production.  As long as $T_P\gtrsim 7.62$,
$^{18}$O/$^{16}$O in the envelope is completely independent of $T_P$.
This is illustrated by plotting in Fig. \ref{fig:o18endpoints} for
$^{18}$O/$^{16}$O vs.  $^{26}$Al/$^{27}$Al the final compositions
after CBP running through the TDU period.  Compositions were computed
with the numerical model.  One can see that the curves of constant
\mdot\ (dotted lines) are approximately curves of constant final
$^{18}$O/$^{16}$O, while the curves of constant $T_P$ (dashed lines)
are approximately curves of constant $^{26}$Al/$^{27}$Al.  The
departure of the constant-\mdot\ curves from constant
$^{18}$O/$^{16}$O is explained as follows: at low $T_P$ (low
$^{26}$Al/$^{27}$Al), $^{18}$O destruction on a single pass through
the radiative zone is incomplete; at high $T_P$ (high
$^{26}$Al/$^{27}$Al), the situation is complicated by partial $^{16}$O
destruction.

The final envelope compositions are the compositions most likely to be
reflected in presolar grains; the most grain formation should occur
with very high mass loss at the very end of AGB evolution (Table
\ref{tab:stages}).  The circles in Fig.  \ref{fig:o18endpoints} show
the abundances found in oxide grains.  It appears that the model can,
in principle, explain simultaneously the $^{26}$Al and $^{18}$O
contents of all the oxide grains found so far, and with reasonable
values of \mdot\ and $T_P$ (however, see the $^{17}$O trends below).
The largest value of $^{26}$Al/$^{27}$Al indicates a maximum $T_P$ in
the source stars of about $10^{7.73}$ K, while a wide range in \mdot\
is suggested if all the $^{18}$O/$^{16}$O values are attributable to
CBP.  Not all grains formed from 1.5 \ms\ stars.  Moreover, inferred
values of \mdot\ and $T_P$ are sensitive to exactly when CBP occurs
and for how long; these values do provide qualitative insight into the
actual mixing rate and depth.  In any case, $^{18}$O/$^{16}$O
decreases at a rate dependent upon \mdot\ and not $T_P$, so that there
is no strong correlation with $^{26}$Al/$^{27}$Al production except
that $^{26}$Al/$^{27}$Al and $^{18}$O/$^{16}$O may be modified
simultaneously by CBP.  Low \mdot\ and high $T_P$ will produce small
$^{18}$O depletions and abundant $^{26}$Al.  High \mdot\ and high
$T_P$ will produce major $^{18}$O depletion and high $^{26}$Al.  Thus,
a direct correlation between $^{26}$Al/$^{27}$Al and $^{18}$O/$^{16}$O
should not be found, although Nittler et al. (1997) and Choi,
Wasserburg \& Huss (1999) have suggested that it might occur on an
empirical basis.

\begin{figure}
\epsscale{1.0}
\plottwo{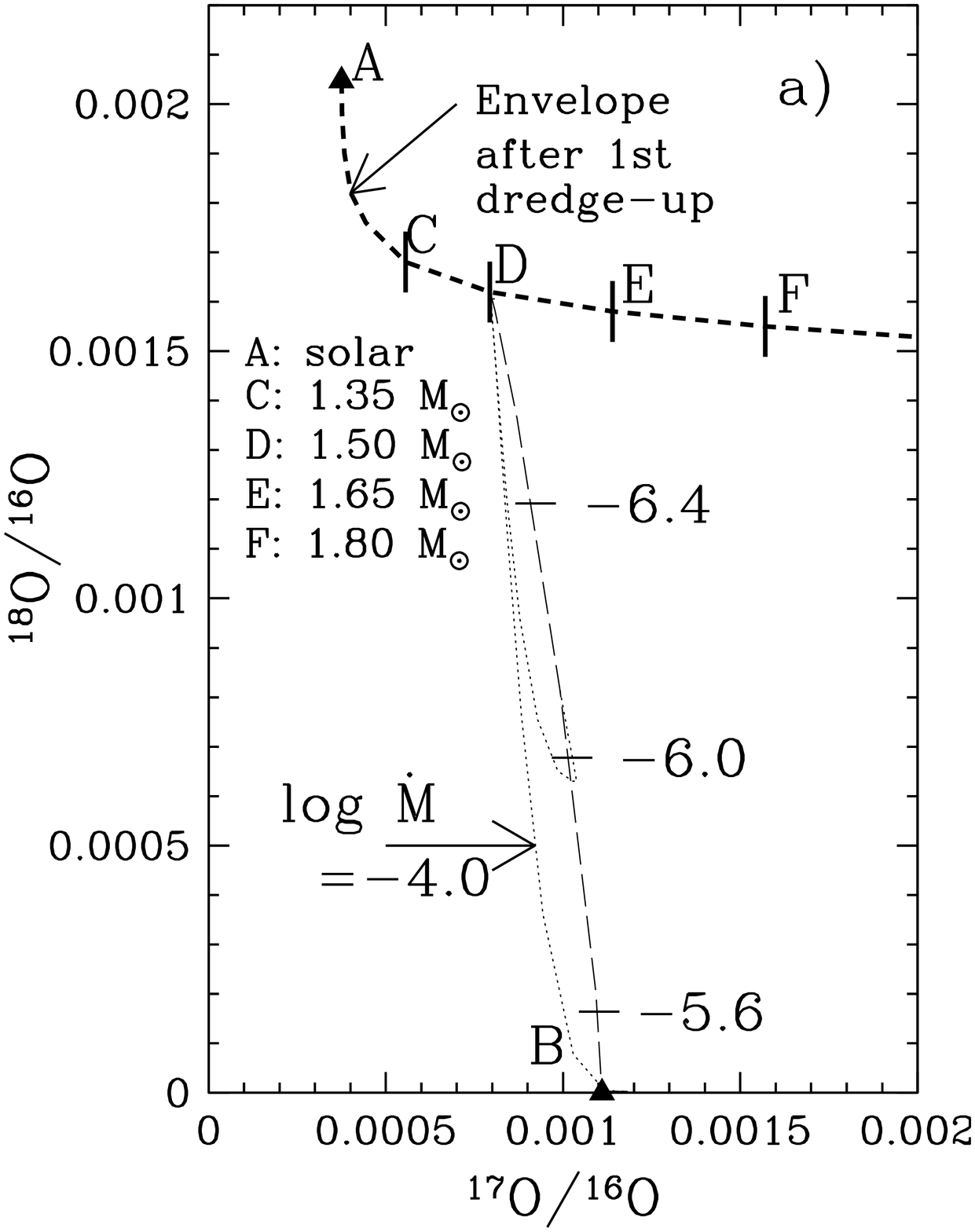}{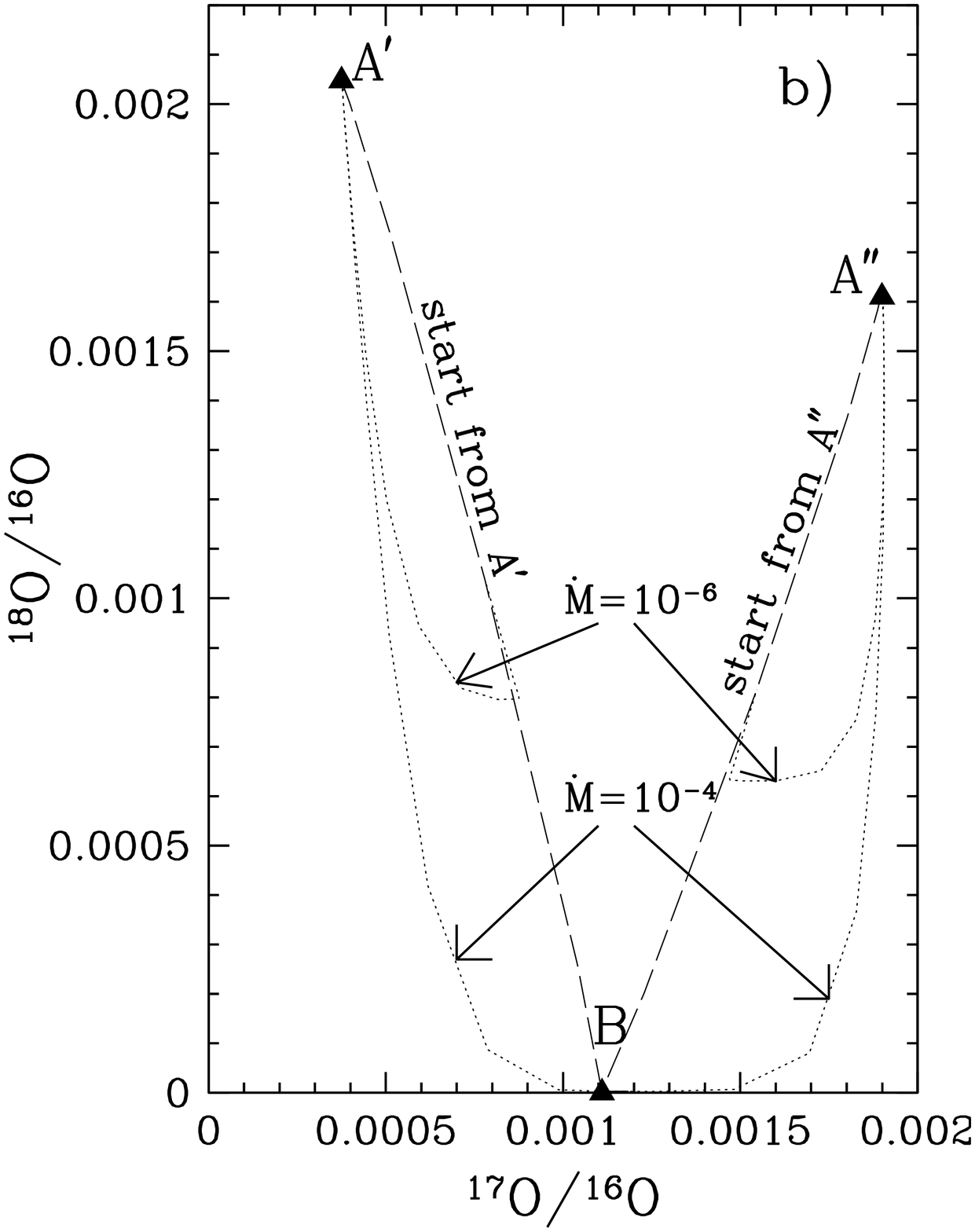}
\epsscale{0.7}
\caption{Final envelope compositions of a 1.5\ms\ star after CBP
($10^{-8}\ms/{\rm yr} \leq \mdot \leq 10^{-4}\ms/{\rm yr}$, $7.44\leq
\log T_P\leq 7.76$, notations as in
Fig. \protect\ref{fig:o18endpoints}).  a) Point D is the composition
after first dredge-up.  The triangle B is the composition reached if
CBP continues for a long time at $\log T_P \ga 7.62$.  The calculated
compositions all lie near a two-component mixing curve connecting
points D and B.  Compositions to the left of this curve indicate
incomplete processing.  The thick dashed curve ACDEF indicates the
results of dredge-up during the red giant phase of stellar evolution
for various initial stellar masses (cf. WBS).  b) trajectories
resulting from CBP for two arbitrary stars with O compositions of
A$^\prime$ and A$^{\prime\prime}$ after first dredge-up.  Mixing lines
point to the equilibrium $^{17}$O/$^{16}$O value.  Range of \mdot\ and
$T_P$ is the same as in panel (a).}
\label{fig:o17endpoints}
\end{figure}

The plateau of $^{17}$O/$^{16}$O at $\mdot>10^{-6}\ms$/yr for all
$T_P$ seen in Fig. \ref{fig:mdotvs17} shows net increases above the
dredge-up value ranging from 0\% to 40\%.  This plateau value
increases with temperature, but is at 0.0011 for all $T_P\geq 7.62$.
Fig. \ref{fig:o17endpoints}a shows $^{18}$O/$^{16}$O
vs. $^{17}$O/$^{18}$O for the final envelope composition, at various
values of \mdot\ and $T_P$.  Generally speaking, the results lie in
the space between point D (composition after first dredge-up) and the
end point B.  If we use Stage 3 instead of Stage 5 to specify the
stellar structure, this changes the results by shifting the
$^{17}$O/$^{16}$O end point B to 0.0016.  Results for Stages 4 and 6
are similar to those of Stage 5.

We now examine $^{17}$O/$^{16}$O using the semi-analytic treatment.
In the circulating material, $^{16}$O destruction is very small, and
the $^{17}$O abundance is governed by competing production via
$^{16}{\rm O}(p,\gamma)^{17}{\rm F}(\beta^+\nu)^{17}{\rm O}$ and
destruction via $^{17}{\rm O}(p,\alpha)^{14}{\rm N}$:
\begin{equation}
\label{eqn:o17burn}
\dot{Y}_{17}^C = Y_{16}^C \lambda[^{16}{\rm O}(p,\gamma)]
                 -Y_{17}^C \lambda[^{17}{\rm O}(p,\alpha)].
\end{equation}
Starting from a low initial abundance, $Y_{17}^C$ increases as matter
moves through the radiative zone until the two terms on the right-hand
side of Eqn. \ref{eqn:o17burn} cancel.  The equilibrium ratio
specified by this cancellation is determined completely by reaction
rates so that processed material has
\begin{equation}
\label{eqn:o17equilibrium}
\frac{Y_{17}^P}{Y_{16}^P} = \frac{\lambda[^{16}{\rm O}(p,\gamma)]}{
                 \lambda[^{17}{\rm O}(p,\alpha)]} \sim 0.0011.
\end{equation}
As a result,
\begin{equation}
\frac{Y_{17}^E}{Y_{16}^E} \sim \frac{Y_{17}^P}{Y_{16}^P}
                              -\left[\frac{Y_{17}^P}{Y_{16}^P}
                              -\left(\frac{Y_{17}^E(0)}{Y_{16}^E(0)}\right)
			      \right]e^{-\mdot t/M_E}
\label{eqn:mixing}
\end{equation}
The final envelope composition using this approximation is shown as a
function of \mdot\ in the dashed curve in Fig. \ref{fig:mdotvs17}.  As
with the numerical results presented, this assumes the radiative zone
of Stage 5 and $t=8.5\times 10^5$ yr.  We chose $M_E=0.5\ms$ for the
analytic approximation.  It is clear that at the condition of $\log
T_P> 7.62$ required to obtain equilibrium, the results are in close
agreement with the numerical calculation.  Thus, the envelope
$^{17}$O/$^{16}$O as a function of time reflects the replacement of
envelope matter with material that has $^{17}$O/$^{16}$O$\sim 0.0011$.

\begin{figure}
\plotone{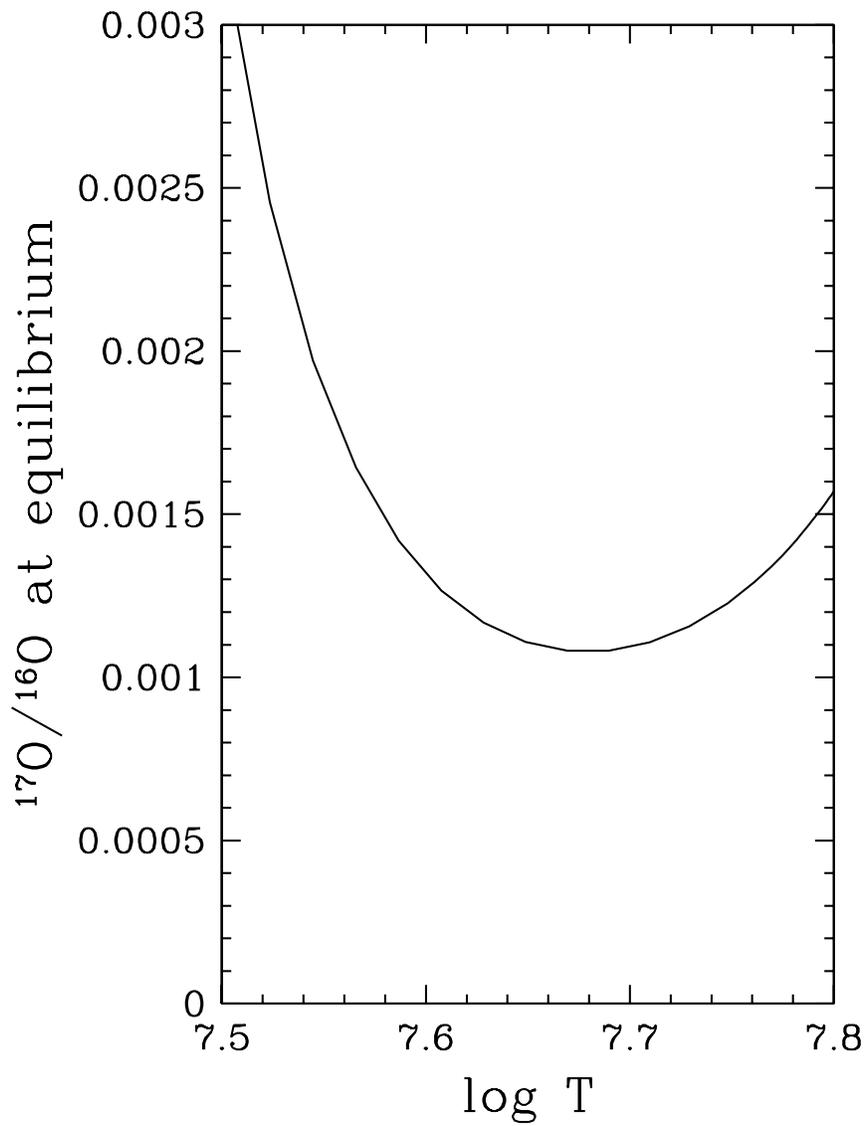}
\caption{Equilibrium $^{17}$O/$^{16}$O abundance ratio as a function
of temperature, which depends only on the reaction rates. The high
values on the left side of the graph are inaccessible in CBP because
the reactions are too slow to establish equilibrium at $\log T_P \la
7.6$ (See Table \protect\ref{tab:stages}).}
\label{fig:17o16eq}
\end{figure}

The steady-state $^{17}$O/$^{16}$O is almost temperature-independent
over the temperatures characteristic of the lower part of the
radiative zone late on the AGB, as shown in Fig. \ref{fig:17o16eq}.
As a result, $Y_{17}^P/Y_{16}^P$ is approximately 0.0011 for all cases
with $\log T_P > 7.62$, with an uncertainty of about 30\% in the ratio
arising from measurement errors on the cross section inputs (Blackmon
et al. 1995; Angulo et al. 1999; see also Appendix \ref{sec:rates}).
We note that only in the last six years have $^{17}{\rm
O}(p,\alpha)^{14}{\rm N}$ cross sections been sufficiently
well-measured to allow us to reach this conclusion, because of the
measurement of the 66 keV resonance by Blackmon et al. (1995).  At
lower temperatures, the equilibrium $^{17}$O/$^{16}$O increases, but
the timescale to reach equilibrium also increases so that these higher
values of $^{17}$O/$^{16}$O are inaccessible after Stage 4.  The
highest value that we found at any time on the TP-AGB is 0.0016.

Since the envelope $^{17}$O/$^{16}$O as a function of time reflects
gradual replacement of envelope material with material that has fixed
$^{17}$O/$^{16}$O = 0.0011, and the processed material also has
extremely low $^{18}$O/$^{16}$O, we now recognize the numerical
results in Fig. \ref{fig:o17endpoints}a as mostly falling on a
two-component mixing curve.

We also carried out calculations using radiative-zone structures from
other stages, and the results were almost indistiguishable as long as
$\log T_P \gtrsim 7.60$.  As explained in the analytical treatment,
the dominant effect for $\log T_P > 7.62$ is to produce an envelope
that is a mixture of the envelope composition after first dredge-up
and the end point B that represents the equilibrium value.  For all
values of $\log T_P\gtrsim 7.62$, the numerical results lie on the
mixing line at positions depending almost exclusively on $\mdot$ as
indicated in the figure.  For lower temperatures, ($\log T_P < 7.62$)
the final compositions fall in a space bounded on the one side by the
mixing curve and on the other by the curve with $\mdot=10^{-4}\ms$/yr.
More generally, as the oxygen isotopic evolution resulting from CBP
depends on the value of $^{18}$O/$^{16}$O and $^{17}$O/$^{16}$O after
first dredge-up and not on the detailed stellar evolution, we may
readily infer the results of CBP for low-mass AGB stars of generally
solar chemical abundances but with arbitrary oxygen isotopic
composition: the end point B is the natural final composition for
material that has been near the H shell in any AGB star of low mass
($M\lesssim 3\ms$), because all of these stars have H shell
temperatures near $10^{7.8}$ K.

The results for two initial compositions are shown in
Fig. \ref{fig:o17endpoints}b.  The upper point $A^{\prime}$ represents
ratios in the envelope at the end of first dredge-up for some star.
For $\log T_P \gtrsim 7.62$, the trajectory is a mixing line between
$A^{\prime}$ and B, where the position only depends on $\mdot t$. If
$\log T_P$ is substantially less than $7.62$, but greater than $7.53$,
it will be a curve that drops from $A^{\prime}$ (due to $^{18}$O
destruction) and then swings toward $B$ as $\mdot$ increases. This
same rule applies to $A^{\prime\prime}$ or any other point, insofar as
the initial stellar mass does not exceed 4--5\ms, where HBB may occur
(Iben \& Renzini 1983). As the initial oxygen isotopic composition of
the star which produced a particular oxide grain is not in general
known, we will use this construction in discussing the data on grains
in Section \ref{sec:oxides}.

In light of the above considerations, the production of
$^{17}$O/$^{16}$O$> 0.0011$ by first dredge-up is explained because
material in the stellar mantle is processed during the main-sequence
phase at much lower temperatures than those present on the AGB.  The
relevant reaction rates are much slower and the equilibrium ratio
larger (see Fig. \ref{fig:17o16eq}), while the duration of the
main-sequence phase is much longer than that of the AGB phase.

\subsection{Carbon}
\label{sec:carbon}

\begin{figure}
\epsscale{1.0}
\plottwo{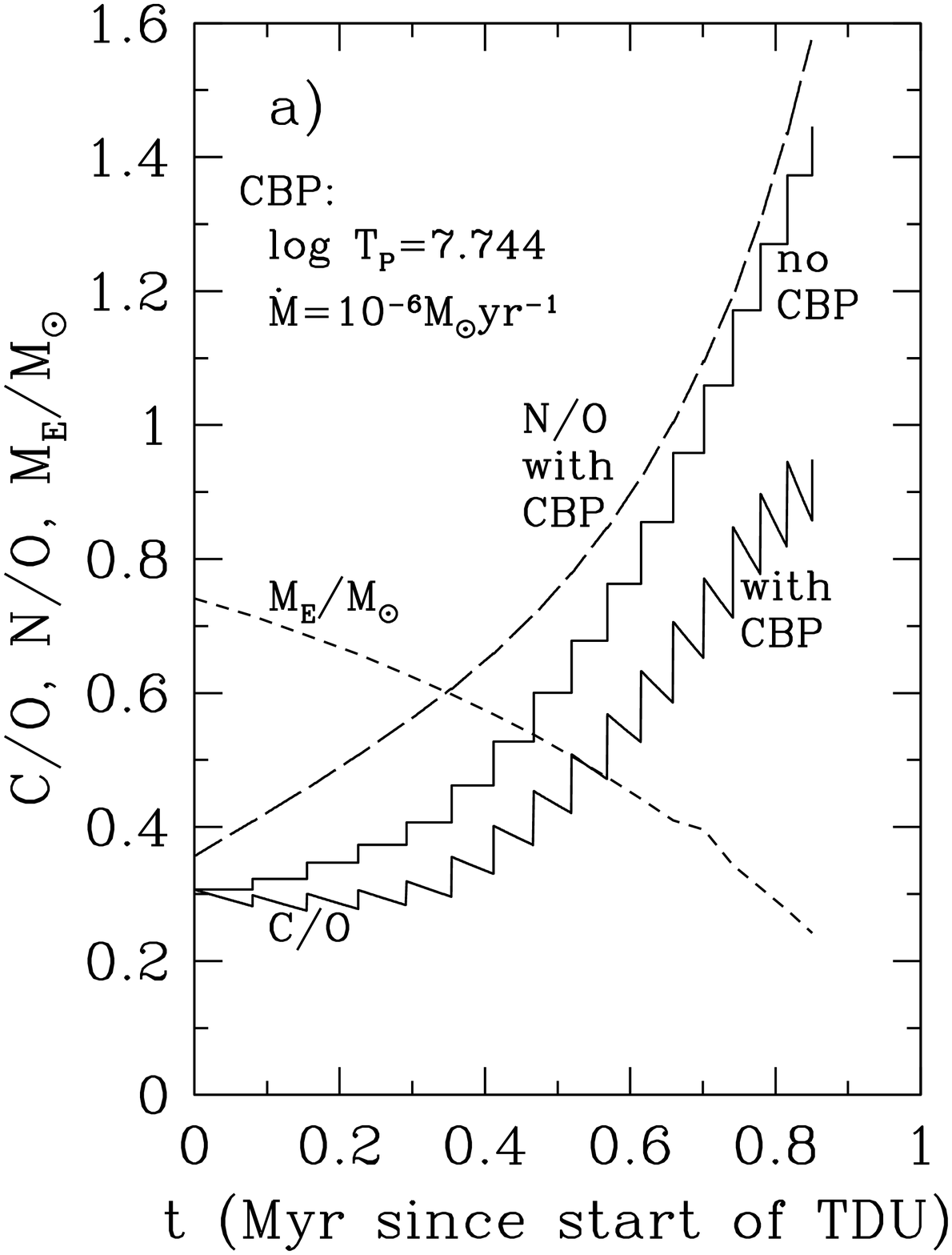}{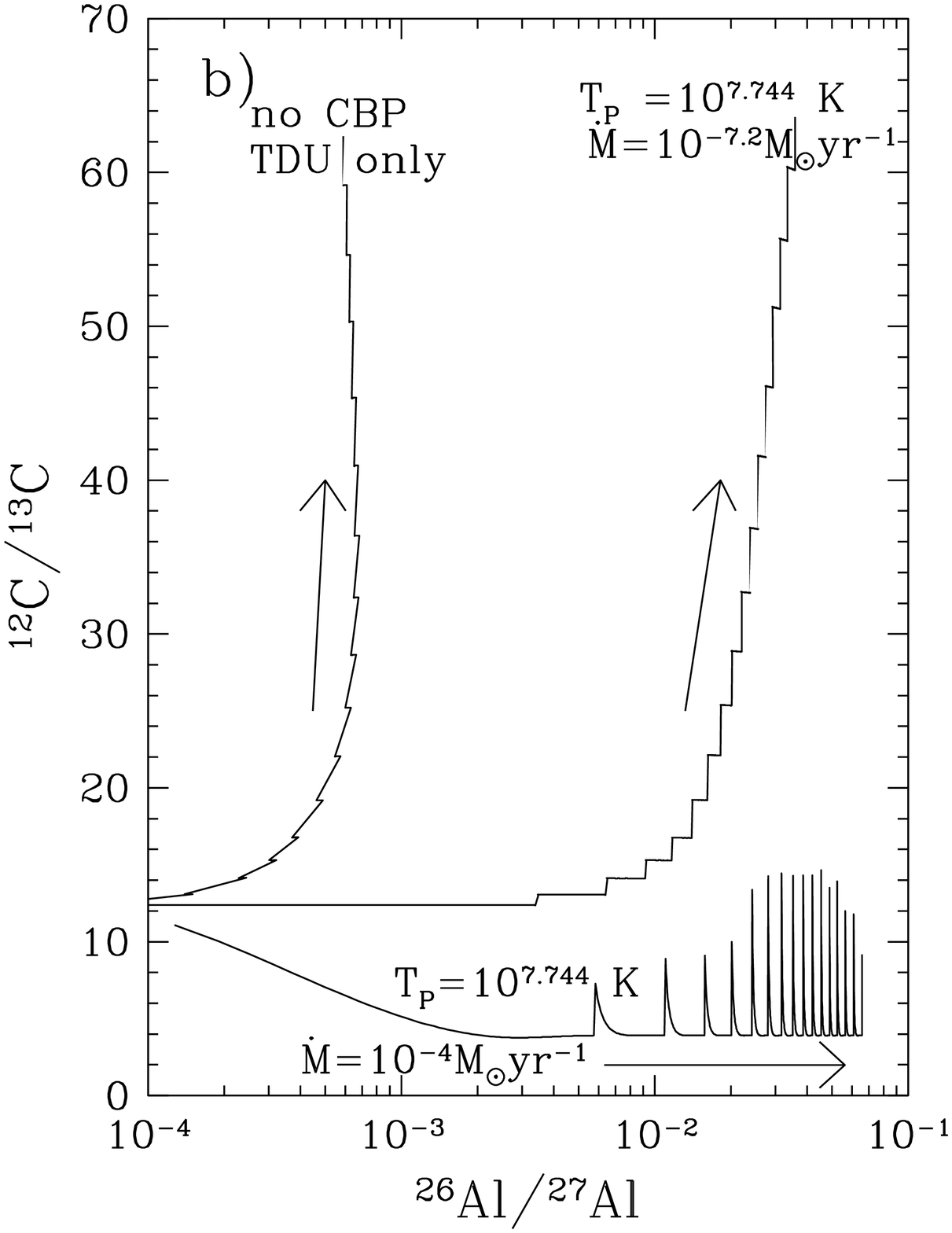}
\epsscale{0.7}
\caption{a) Time evolution of C/O in SEM(1.5) (no CBP).  Also shown
is the envelope mass as a function of time.  The C/O curve labelled
``with CBP'' is the evolution of C/O with time for a star with CBP.
The N/O curve is for the same CBP model.  Note that CBP adds to N/O in
the envelope at the expense of C/O.  b) Evolution of $^{12}$C/$^{13}$C
and $^{26}$Al/$^{27}$Al with time in the numerical model of concurrent
CBP and TDU, at two choices of \mdot\ and compared with the no-CBP
model.  Arrows indicate the general direction of evolution.  The jumps
in composition are due to dredge-up episodes, and motion toward lower
$^{26}$Al reflects radioactive decay.  Increasing $T_P$ moves the
end points of these curves to the right.  Increasing \mdot\ moves the
end points downward.  }
\label{fig:tdu-cartoon}
\end{figure}

The evolution of carbon in CBP is more complex than that of $^{26}$Al,
$^{17}$O, or $^{18}$O, since the nuclides discussed so far are
negligibly affected by TDU.  The abundance of $^{12}$C in the envelope
is altered at each dredge-up episode.  Consequently, the
$^{13}$C/$^{12}$C ratio is shifted.  The change in C/O in the envelope
at each thermal pulse in SEM(1.5) is shown in Figure
\ref{fig:tdu-cartoon}a along with the change in the mass of the
envelope with time. In the numerical model, we assume that CBP starts
immediately after the first thermal pulse that produces dredge-up
(Table \ref{tab:stages} and Fig. \ref{fig:tpagb}).  This is the same
calculation used for $^{26}$Al, $^{18}$O, and $^{17}$O.

\begin{figure}
\plotone{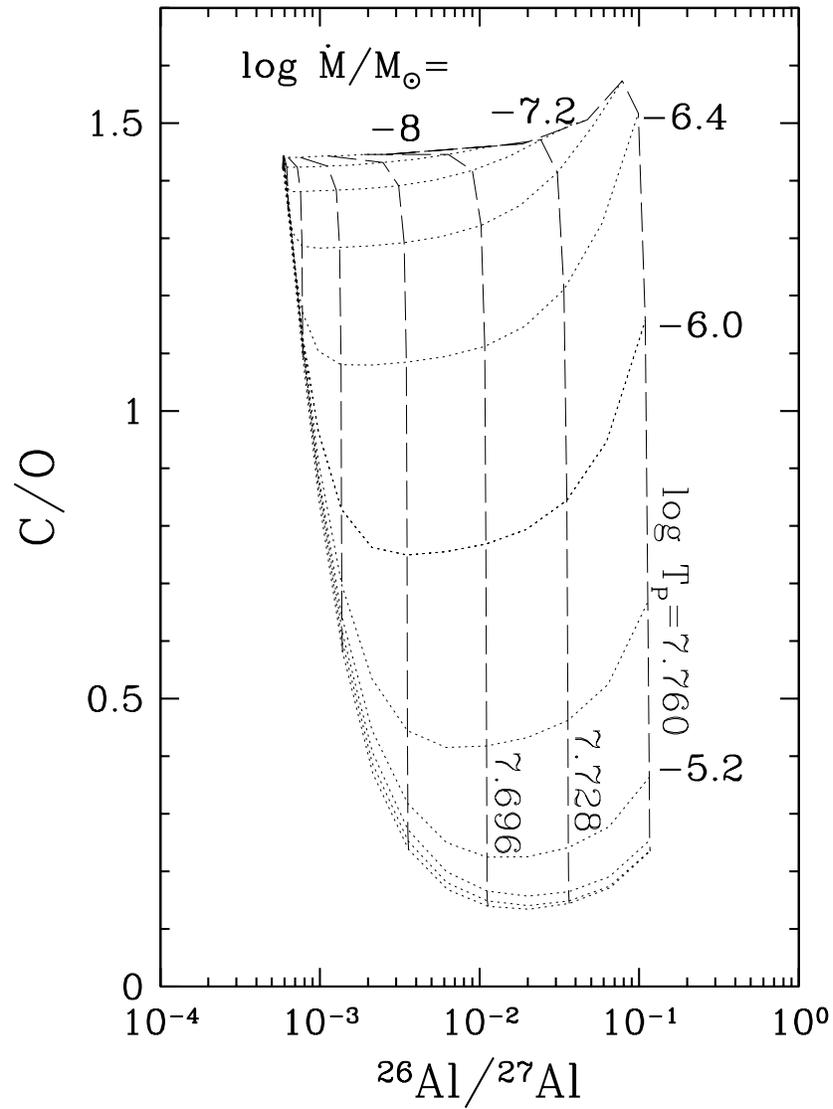}
\caption{Final envelope compositions, with $7.44 \leq \log T_P \leq
7.76$ and $10^{-8}\ms/{\rm yr} \leq \mdot \leq 10^{-4}\ms/{\rm yr}$.
Note that a carbon star is produced when C/O$> 1$, corresponding in
most cases to $\mdot< 10^{-6}\ms$/yr.}
\label{fig:co}
\end{figure}

\begin{figure}
\epsscale{0.6}
\plotone{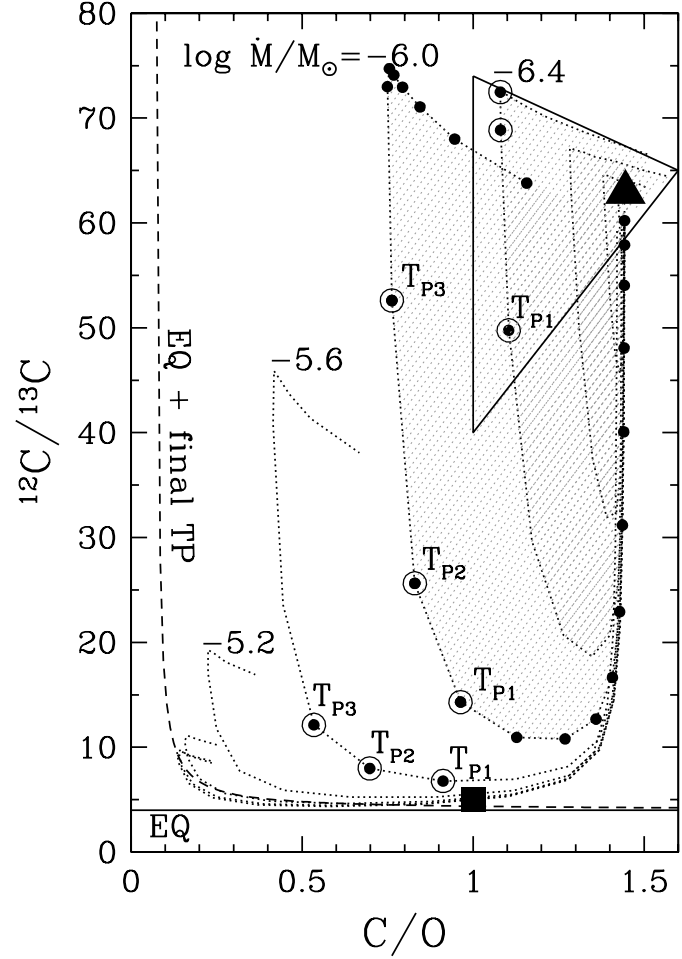}
\caption{Final compositions of the envelope where CBP stops
immediately before the last thermal pulse, after which the last
dredge-up episode adds $^{12}$C.  $T_P$ increases clockwise along the
constant-\mdot\ curves; dots on the $\mdot=10^{-6}\ms$/yr curve
indicates $\log T_P$ values in steps of 0.016, and points with $\log
T_{P1}=7.616$, $\log T_{P1}=7.632$, and $\log T_{P3}=7.648$ are
circled and labelled.  The solid triangle indicates the final
composition of a 1.5\ms\ star without CBP.  Shaded region contains
$\mdot < 10^{-6}\ms$/yr, approximately the region of carbon star
formation, while the large open triangle encloses compositions which
have C/O$>1$ and $^{26}$Al/$^{27}$Al$> 10^{-3}$.  Note that
low-$^{12}$C/$^{13}$C carbon stars are restricted to a small region in
the lower right-hand corner, near the solid square.  EQ is the
limiting equilibrium value.  Bounding curve ``EQ + final TP''
corresponds to $^{12}$C added by the final dredge-up episode, mixing
with an envelope of composition EQ.}
\label{fig:13vCOtdu}
\end{figure}

Figure \ref{fig:tdu-cartoon}a shows the evolution of C/O for SEM(1.5)
as a function of time up until the end of TDU and includes the cases
of CBP with $\mdot = 10^{-6.0}\ms$/yr and $\log T_P = 7.696$ for
comparison. Note that C/O = ($^{12}$C + $^{13}$C)/$^{16}$O. In Figure
\ref{fig:tdu-cartoon}b we show the evolution of $^{12}$C/$^{13}$C in
the envelope versus $^{26}$Al/$^{27}$Al with time for three cases.  As
already stated, it is evident that high values of $T_P$ give high
$^{26}$Al/$^{27}$Al. However, to obtain low $^{12}$C/$^{13}$C at high
$^{26}$Al/$^{27}$Al requires high $\mdot$ (e.g. $10^{-5} \ms$/yr).
In the latter case, the $^{12}$C content of the envelope is so low
that every thermal pulse gives a spike in $^{12}$C/$^{13}$C which is
then destroyed by rapid CBP. The final envelope compositions in terms
of C/O and $^{26}$Al/$^{27}$Al are shown in Figure \ref{fig:co}.  It
can be seen that the approximate boundary for C/O $\gtrsim 1$ is
$\mdot \approx 10^{-6.4}\ms$/yr.  The relationship between
$^{12}$C/$^{13}$C and C/O is shown in Figure \ref{fig:13vCOtdu}.  The
solid triangle represents the final envelope composition of the AGB
star computed by SEM(1.5) with no CBP.  The results for CBP are shown
as curves of constant $\mdot$ with the values of $\log T_P$ varying
with position along these curves.  It can be seen that to obtain both
C/O $> 1$ and low $^{12}$C/$^{13}$C ( $\lesssim 20$) requires
conditions that are very restrictive ( $\log T_P \lesssim 7.616$).
Low $^{12}$C/$^{13}$C may readily occur for higher $T_P$ but with C/O
$< 1$.

In the material circulating through the radiative zone, the $^{12}$C
and $^{13}$C abundances are governed by the equations
\begin{eqnarray}
\dot{Y}_{12}^C & = & Y_{15}^C\lambda[^{15}{\rm N}(p,\alpha)]
                     -Y_{12}^C\lambda[^{12}{\rm C}(p,\gamma)]
\label{eqn:c12destruction}\\
\dot{Y}_{13}^C & = & Y_{12}^C\lambda[^{12}{\rm C}(p,\gamma)]
                     -Y_{13}^C\lambda[^{13}{\rm C}(p,\gamma)].
\label{eqn:c13production}
\end{eqnarray}
As a sample of circulating material descends into the radiative zone
and its temperature rises, the negative term on the right-hand side of
Eqn. \ref{eqn:c12destruction} grows and $^{12}$C is destroyed.  The
$^{12}$C that is destroyed is converted to $^{13}$C, so that quickly
both the production and destruction terms in
Eqn. \ref{eqn:c13production} balance.  The condition that
$^{12}$C/$^{13}$C in this material does not change with time then
gives the steady-state ratio
\begin{equation}
\frac{Y_{12}^C}{Y_{13}^C}=\frac{\lambda[^{13}{\rm C}(p,\gamma)]
                                -\lambda[^{12}{\rm C}(p,\gamma)]}{
                                 \lambda[^{12}{\rm C}(p,\gamma)]},
\end{equation}
which depends only on cross sections.  For the temperatures in the
radiative zone, it is about 3.  As processing continues, $^{12}$C
continues to burn while $^{12}$C/$^{13}$C remains approximately
constant, producing $^{14}$N.  Finally, if $T_P$ is sufficiently high,
enough $^{14}$N is made to close the CN cycle with the reaction chain
$^{14}\mathrm{N}(p,\gamma)^{15}\mathrm{O}
(\beta^+\nu)^{15}\mathrm{N}(p,\alpha)^{12}\mathrm{C}$, and the amounts
of $^{12}$C and $^{13}$C become steady with $^{12}$C/$^{13}$C$\sim 4$
and $Y_{12}^C\sim 0.03 [Y^E_{12}(t)+Y^E_{13}(t)]$.  Thus, CBP destroys
$^{12}$C, mixing $^{12}$C-depleted and $^{13}$C-enhanced material into
the envelope.  This explains the strong dependence of the C isotopes
on \mdot: because at most temperatures, \mdot\ sets the rate at which
processed material of the new $^{12}$C/$^{13}$C composition is mixed
into the envelope.

The effects of CBP can be seen as follows for a given \mdot.  The
ratio $^{12}$C/$^{13}$C first decreases with increasing $T_P$ due to
increasing input of processed material, approaching
$^{12}$C/$^{13}$C$\sim 3$.  For $\log T_P\gtrsim 7.6$, most of the C
is converted to $^{14}$N in the processed material, so that processed
material contains only small amounts of $^{12}$C and $^{13}$C.  As a
consequence, $^{12}$C/$^{13}$C in the envelope remains near the value
it would have without CBP.  For $\mdot\gtrsim 10^{-6.4}\ms$/yr, the
processing at $\log T_P\gtrsim 7.6$ is so effective that a substantial
amount of carbon is converted to N and C/O becomes less than one.

\subsection{C-rich vs. O-rich envelopes}
\label{sec:coconditions}

The conversion of a large fraction of the carbon in the stellar
envelope into nitrogen has important observational effects: the C/O
ratio determines whether the signs of related nucleosynthesis should
be expected in S stars versus carbon stars, or in oxide grains versus
carbon and carbide grains.  As seen in Figs. \ref{fig:tdu-cartoon}a
and \ref{fig:co}, when (and whether) this ratio exceeds unity depends
on competition between third dredge-up and CBP in the invervals
between dredge-up events.  To look at this competition, we now examine
the problem analytically.

The rate of addition of $^{12}$C to the envelope by TDU is some
function of time, $F(t)$.  CBP removes material from the envelope and
replaces it with material in which some of the $^{12}$C has been
destroyed.  The time evolution of the envelope $^{12}$C abundance is
then governed by the equation
\begin{equation}
\dot{Y}_{12}^E = \frac{\mdot}{M_E}\left(Y_{12}^P - Y_{12}^E\right)
+ M_E^{-1}F(t).
\end{equation}
We wish to find the condition on \mdot\ such that the envelope is
carbon-rich at the end of the TP-AGB.  If
$Y^P_{12}=(1-\alpha)Y^E_{12}$ (where $\alpha$ is the fraction of
$^{12}$C destroyed in a single pass), we then have
\begin{equation}
\dot{Y}_{12}^E=-\frac{\mdot}{M_E}\alpha Y_{12}^E + F/M_E.
\label{eqn:codiffeq}
\end{equation}
$F(t)$ is essentially a sum of delta functions from dredge-up, but we
will approximate it by its mean value, $\langle F\rangle =12\times
4.3\times 10^{-6} Y_{12}^E(0)M_E\ {\rm yr}^{-1}$ (see
Fig. \ref{fig:tdu-cartoon}a).  The solution to Eqn. \ref{eqn:codiffeq}
is
\begin{equation}
Y_{12}^E(t)=e^{-\mdot\alpha t/M_E}Y_{12}^E(0) + \frac{\langle
F\rangle}{\mdot\alpha}\left(1-e^{-\mdot\alpha t/M_E}\right).
\label{eqn:cocondition}
\end{equation}
At the end of the TDU phase, $t=8.5\times 10^5$ yr.  The condition
that the envelope is carbon-rich is that $Y_{12}^E/Y_{16}^E > 1$.
Inserting this condition along with $Y_{12}^E(0)=1.7\times 10^{-4}$
into Eqn. \ref{eqn:cocondition}, dividing by $Y_{16}^E=5.9\times
10^{-4}$, and solving for \mdot\ gives
\begin{equation}
\mdot < 7\times M_E \times 10^{-7}\ {\rm yr}^{-1}/\alpha.
\label{eqn:conumerical}
\end{equation}
The value of the fraction destroyed ($\alpha$) depends strongly on
$T_P$ and weakly on \mdot.  A reasonable mean value of $M_E$ during
the TDU period is $0.4 \ms$.  If $\alpha\sim 1$, then $\mdot \la
3\times 10^{-7}\ms$/yr for C/O $> 1$, within a factor of a few of what
we find numerically.  If $\mdot=10^{-4}\ms$/yr, the maximum CBP rate
we considered, then C/O$>1$ in the envelope requires $\alpha\leq
3\times 10^{-3}$, or very little carbon destruction per pass.  The
combination of high \mdot\ and C/O$>1$ is therefore restricted to very
low $T_P$.

If the true efficiency of dredge-up is different from that for
SEM(1.5), then $\langle F\rangle$ differs from the value used here
(for a possible mechanism, see Herwig et al. 1997).  If $\langle
F\rangle$ is doubled, then the \mdot\ that divides C/O $>1$ from
C/O$<1$ (Eqn. \ref{eqn:conumerical}) increases by a factor of three;
we find that the relationship between the two parameters is
approximately linear over this range of $\langle F\rangle$.  Since
dredge-up does not have a strong effect on the other abundances, the
only influence on our results is to change which \mdot\ lie on which
side of C/O$<1$.  The abundance changes found above for the O and
below for the N isotopes take place over wider ranges in \mdot\ than
factors of three, so we do not expect qualitative changes in our
results if the dredged-up compositions change.

\subsection{Nitrogen}
\label{sec:nitrogen}

The effect of CBP on nitrogen is to efficiently destroy $^{15}$N and
to greatly increase $^{14}$N in the envelope.  The increase in
$^{14}$N is due to the conversion of $^{12}$C into $^{14}$N by CBP at
$\log T_P \gtrsim 7.55$.  Complete conversion of $^{12}$C in the
envelope from the initial C inventory in the absence of TDU would
increase the $^{14}$N abundance to over three times the original main
sequence value.  As TDU considerably increases the amount of available
carbon in the envelope, even more $^{14}$N will be produced. This can,
at the end of the stellar lifetime, give $1.5 \leq {\rm N/O} \lesssim
8.6$ (with values above 1.7 involving destruction of oxygen).  Such
large amounts could constitute an important source of primary (scaling
linearly with metallicity) nitrogen in the Galaxy.

The relationship between $^{15}$N/$^{14}$N and $^{26}$Al/$^{27}$Al is
shown in Fig. \ref{fig:n15}a for the final envelope composition as a
function of $T_P$ and \mdot\ for the numerical model.  It is seen that
there is a rapid drop in $^{15}$N/$^{14}$N with increasing \mdot\ for
all $\log T_P>7.50$.  For $\log T_P> 7.62$, the $^{15}$N/$^{14}$N
ratio is essentially constant for a given \mdot.  Comparing with
Fig. \ref{fig:co}, we see that a wide range in $^{15}$N/$^{14}$N (from
about $3\times 10^{-5}$ to $6\times 10^{-4}$) may occur if C/O$>1$.
If C/O$<1$ then $\mdot \gtrsim 10^{-6.4}\ms$/yr and and
$^{15}$N/$^{14}$N$\lesssim 3\times 10^{-4}$.  See the discussion of
SiC grains in Section \ref{sec:carbides}.

In Fig. \ref{fig:n15}b we show N/O vs. C/O for the final composition
of the envelope in the numerical model.  The curves for different
\mdot\ and $T_P$ are indicated.  It can be seen that CBP will produce
high N/O for $\mdot\gtrsim 10^{-6.4}\ms$/yr and $\log T_P\gtrsim 7.6$.
We note that C/O$\geq 1$ requires $\mdot \lesssim 10^{-6.4}\ms$/yr for
the same $T_P$.

\section{Self-consistency and relation to other models}
\label{sec:consistency}

\subsection{Effects on energy generation} 

We now examine the self-consistency of the post-processing approach
with the basic AGB stellar model.  Since we assume an underlying
stellar model [SEM(1.5)] without CBP, it is important that the
additional processing not affect the energy generation in a way that
would significantly alter the evolution.  The important issue is to
determine what values of $T_P$ and \mdot\ plausibly allow a
self-consistent calculation.

\begin{figure}
\plotone{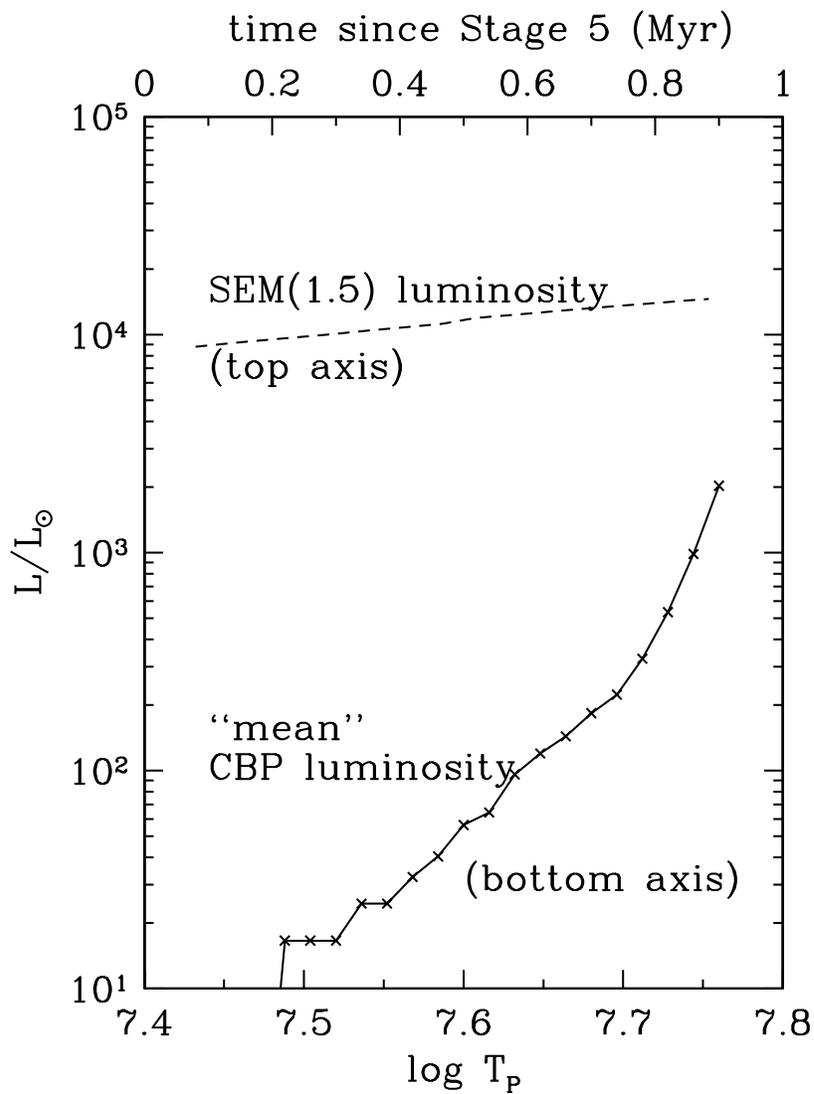}
\caption{Solid curve: mean luminosity in the CBP circulating regions
as a function of $T_P$, estimated as described in the text (read
bottom axis).  Dashed curve: luminosity of the SEM(1.5) model
without CBP, as a function of time (read top axis).}
\label{fig:cbplum}
\end{figure}

For \mdot, we have considered the range $10^{-8}$ to $10^{-4}$\ms/yr.
The low end is comparable to the rate of advance of the H shell, and
too slow to turn over the radiative region between two successive
thermal pulses.  At the high end, the rate at which the radiative zone
turns over corresponds to $\mdot/\Delta M\sim 0.1$ times per year.  If
the CBP circulation is to be distinct from envelope convection, its
turnover rate should be at least an order of magnitude less than that
of the convective envelope ($\sim 1$ per year).  Concerning the
effective penetration depth $T_P$, we chose the lower limit of 7.44
for negligible processing and the upper limit of 7.76 to produce
$^{26}$Al/$^{27}$Al$\sim 10^{-1}$.  With this range in parameters we
now examine the effects on the energy production rate and luminosity.
The energy generation in the parts of the stellar interior where CBP
takes place occurs at very nearly the same rate with and without CBP.
This is because CBP never changes the H mass fraction on a single pass
through the radiative zone by more than $10^{-4}$ in the CBP models.
It was calculated that CBP will change the H mass fraction in the
envelope over the whole evolution by a total of only 4.6\% even at the
highest $T_P$ we considered.  Since we have assumed that CBP changes
neither the physical conditions in the radiative zone nor the total
number of C and N catalyst nuclei, the energy generation rate can
change by at most 4.6\%.  An estimate of the energy generated in the
matter undergoing CBP is computed as follows: we examine the H mass
fraction of the envelope at the end of the numerical calculation and
find the difference from the final mass fraction in the original
SEM(1.5) model in the absence of CBP.  We multiply this difference
by an average envelope mass of 0.6\ms\ and then by 7 MeV/nucleon, and
divide by the $8.5\times 10^5$ yrs that CBP was active.  This yields
an effective luminosity, shown in Fig. \ref{fig:cbplum} as a function
of $T_P$.  For $\mdot> 10^{-7}$\ms/yr, this luminosity is independent
of \mdot\ (because, just as in the case of $^{26}$Al, only a small
amount of processing occurs on each passage through the radiative
zone).  Also plotted in Fig. \ref{fig:cbplum} is the evolving
luminosity of the SEM(1.5) model for comparison.

For $\log T_P=7.76$, the CBP flow yields at most 25\% of the star's
total energy generation rate during the TP-AGB phase.  However, this
amount is very close to the energy generation that would occur in the
same radiative layers without CBP; the difference is 1.3\% of the
star's total energy output.  Our proposal to explain
$^{26}$Al/$^{27}$Al ratios by CBP requires $\log T_P$ extending only
to $\log T_P\sim 7.73$; this corresponds to 5\% of the total energy
generation occurring in the CBP flow and depletes about 1.2\% of the
envelope hydrogen.  It follows that the CBP necessary to produce
$^{26}$Al/$^{27}$Al$\sim 10^{-2}$ as found in some grains does not
have a major effect on the luminosity or the energy generation rate.

\subsection{Extra mixing by diffusion parametrization}

\begin{figure}
\plotone{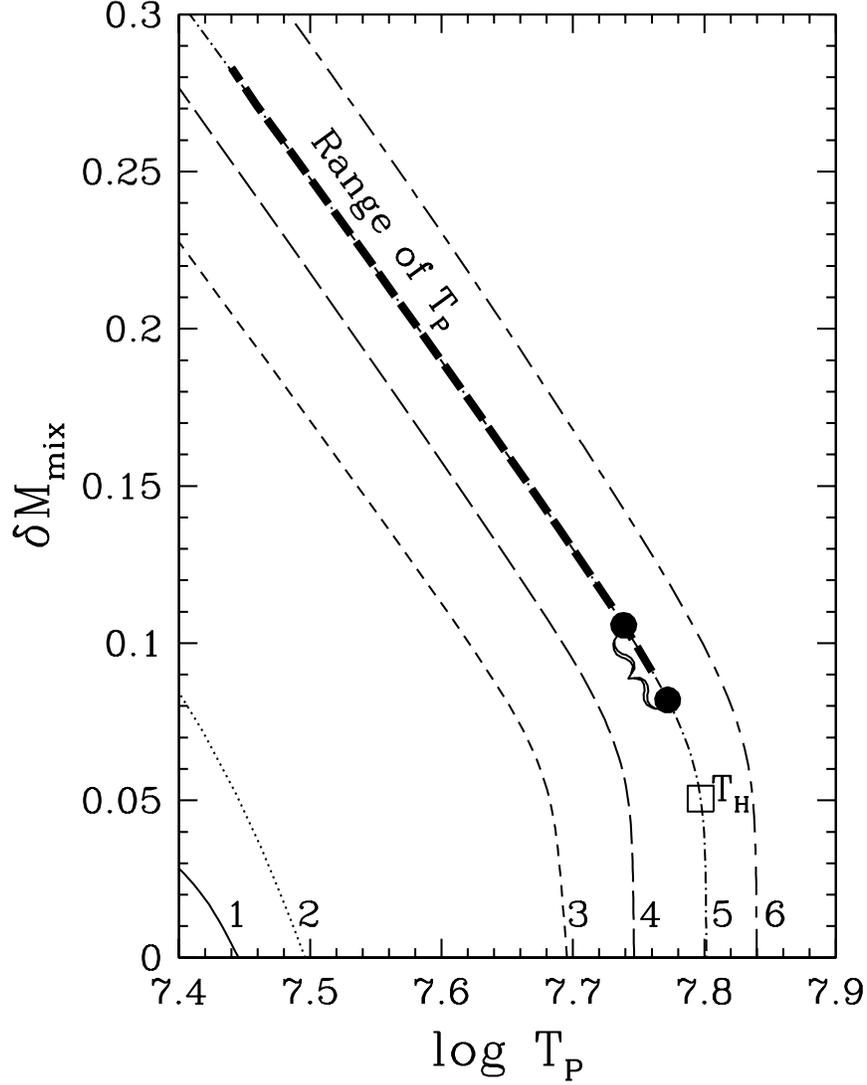}
\caption{Relation between the parameters $\delta M_\mathrm{mix}$ and
$T_P$, at each of the 6 selected evolutionary stages of the radiative
zone (cf. Table \protect\ref{tab:stages} and
Fig. \protect\ref{fig:integral}b).  Along the Stage 5 curve, the thick
broken line indicates $T_P$ considered in this study; the connected
black dots indicate depths chosen to have H depleted by 5--20\% as
assumed for $\delta M_\mathrm{mix}$ in Weiss, Denissenkov \&
Charbonnel (2000).  This shows the essential equivalence of the mass
flow approach (\mdot,$T_P$) and the diffusive mixing approach.}
\label{fig:DM}
\end{figure}

We now wish to compare models in which extra mixing is modelled as
diffusive transport with those in which it is modelled as a mass flow
(the present work; Boothroyd, Sackmann \& Wasserburg 1994, 1995; WBS;
Boothroyd \& Sackmann 1999; Sackmann \& Boothroyd 1999; Messenger
2000).  The mass-flow models have only two parameters, $T_P$ and
\mdot.  In the case of diffusive models, the mixing is modelled either
by post-processing calculations which include time evolution of the
radiative zone, or by inclusion in a full stellar evolution code.  In
either case, a term is added to the stellar evolution model for
diffusion between the convective envelope and a region of the
radiative zone.  The parameters used are a diffusion coefficient
$D_\mathrm{mix}$ and a maximum depth of diffusive mixing $\delta
M_\mathrm{mix}$.  The parameter $\delta M_\mathrm{mix}$ is defined in
terms of the deepest mass coordinate penetrated, $M_\mathrm{mix}$, the
convective-envelope boundary $M_\mathrm{BCE}$, and the base of the H
shell $M_\mathrm{H}$ (see Fig. \ref{fig:cartoon}) as
\begin{equation}
\delta M_\mathrm{mix}=
\frac{M_\mathrm{mix}-M_\mathrm{H}}{M_\mathrm{BCE}-M_\mathrm{H}}.
\end{equation}
To obtain adequate nuclear processing, the diffusive penetration must
extend into the upper part of the H shell.  To prevent the catastrophe
of interference with the region of maximum energy generation, a
barrier boundary is assumed across which no diffusion is permitted to
occur: this is specified as $\delta M_\mathrm{mix}$ (cf.  Denissenkov
\& Weiss 1996; Denissenkov et al. 1998; Weiss, Denissenkov \&
Charbonnel 2000).  The two versions of transport -- mass flow and
diffusive mixing -- are thus closely related and have similar effects.
The relationships between $T_P$ and $\delta M_\mathrm{mix}$ and
between \mdot\ and $D_\mathrm{mix}$ may be seen as follows.  The
parameters $T_P$ and $\delta M_\mathrm{mix}$ are directly related
through the temperature at the position where the diffusive mixing is
stopped.  This is fixed by the basic stellar model.  The relationship
between $D_\mathrm{mix}$ and \mdot\ may be seen as follows.  The
length scale $l$ and time scale $t$ for mixing must satisfy
$l^2/Dt\sim 1$, so that $D_\mathrm{mix}\approx lv$.  For mass flow,
the velocity $v$ corresponding to \mdot\ at radial position $r$ in the
star is $v=\mdot/[4\pi r^2 \rho(r)]$, where $\rho(r)$ is the mass
density.  This yields
\begin{eqnarray}
D_\mathrm{mix} & \approx & \frac{l\mdot}{4\pi r^2 \rho}\\
               & \approx & 2 \times 10^8 \mathrm{cm^2/s}\ 
\left(\frac{ \mdot }{10^{-8}\ms\mathrm{/yr}}\right)
\left(\frac{l}{3\times 10^{10}\mathrm{cm}}\right)
\left(\frac{\rho}{7 \mathrm{g/cm^3}}\right)^{-1}.
\end{eqnarray}
We have taken for reference the values of $r$ and $\rho$ at the H
shell and $l$ the thickness of the radiative zone, all at Stage 5.
This is the basic relationship between $D_\mathrm{mix}$ and \mdot\ to
an order of magnitude.  For comparison, the value estimated by
Denissenkov \& Weiss (1996) is $D_\mathrm{mix}\sim
10^8\,\mathrm{cm}^2$/s in their study of a low-metallicity RGB star.
The relations between $\delta M_\mathrm{mix}$ and $T_P$ for the
evolutionary stages that we considered for CBP are shown in
Fig. \ref{fig:DM}.  In the parametric diffusion treatments of
Denissenkov and co-authors, $\delta M_\mathrm{mix}$ was chosen to be
where the H mass fraction was 5--20\% below the envelope value before
the onset of mixing.  We indicate the corresponding points in
Fig. \ref{fig:DM}.  Their prescription results in similar mixing
depths to those that we have examined in this paper.

\subsection{The problem of $^{26}$Al production and $T_P$}

\begin{figure}
\plotone{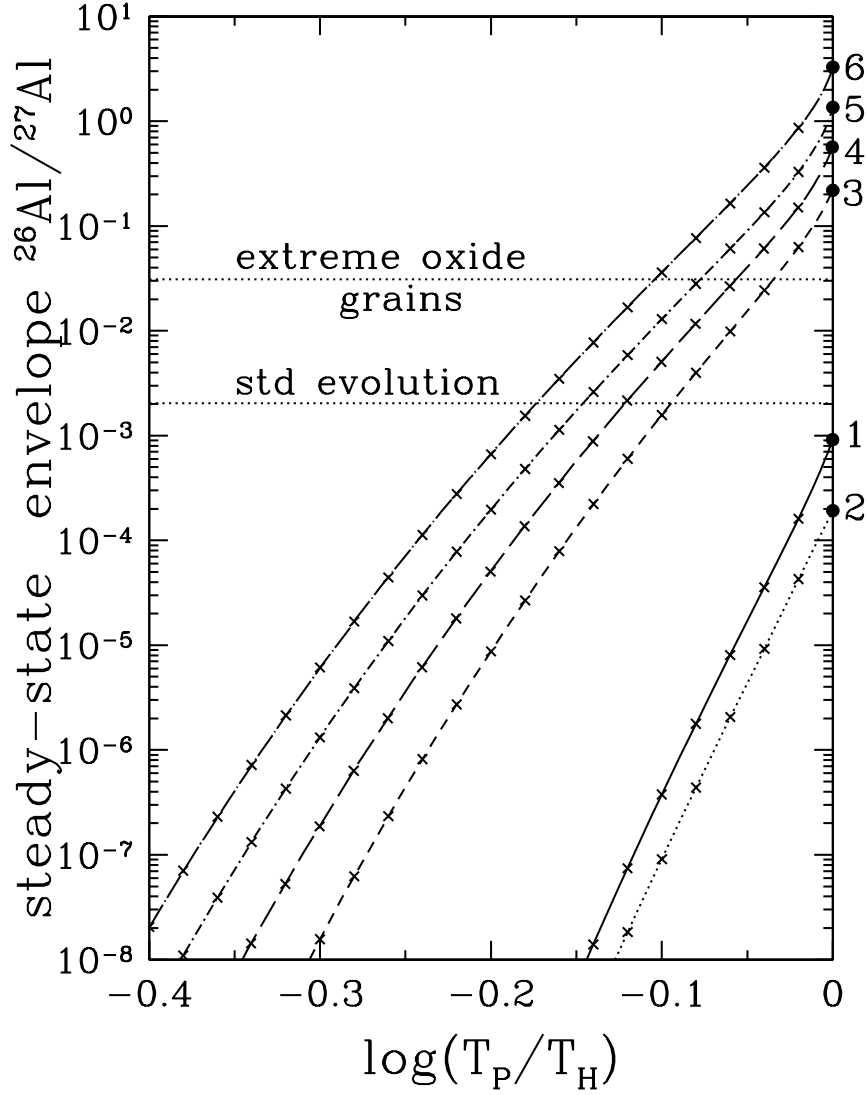}
\caption{Steady-state $^{26}$Al/$^{27}$Al ($\tau_{26}P/Y^E_{27}$)
calculated for the envelope assuming the whole evolution takes place
with the radiative-zone structure fixed at Stages 1--6, as labelled.
The results are as in Fig.  \protect\ref{fig:integral}b but in terms
of $\log\left(T_P/T_H\right)$.  The $^{26}$Al production is governed
by the last stages.  Note that for Stages 4, 5, and 6, abundant
$^{26}$Al is produced for $\log\left(T_P/T_H\right)> -0.10$ and that
$P\propto M_E^{-1}$.}
\label{fig:deltat}
\end{figure}

There remains the issue of whether the calculated effects of CBP for a
star may be reasonably carried out by choosing a given $T_P$ and
\mdot\ and using a selected radiative-zone structure from a single
time (stage) during the TP-AGB.  A complete calculation of CBP would
require inclusion of the detailed evolution with time as the stellar
structure evolves.  Again let us consider a semi-analytic treatment to
understand the results.  The temperature dependence of the processing
is unimportant for carbon, nitrogen, and oxygen isotopes for all
$T_P\gtrsim 7.60$.  The results for these nuclides are thus almost
independent of which stellar structure is chosen.  We verified this by
calculations with the numerical model, utilizing radiative-zone
structures from Stages 3, 4, 5 and 6.  This is not the case for
$^{26}$Al, as can be seen by examining Fig. \ref{fig:integral}b.  For
a given $T_P$, we see that the equilibrium $^{26}$Al/$^{27}$Al for
different times along the evolution (Eqn. \ref{eqn:prodrate-1})
decreases with increasing time.  This rapid decrease occurs because as
the H shell advances outward, the thickness (in mass coordinates) of
the radiative zone decreases sharply, while the ratio of temperatures
at the bottom and top of this region, $T_H/T_{BCE}$ (where $T_{BCE}$
is the temperature at the bottom of the convective envelope), remains
relatively fixed at about 24.  The region in which $^{26}$Al is
produced accordingly becomes thinner very rapidly as the star evolves,
and the amount of $^{26}$Al produced drops.  Another contributing
factor is that the density in the radiative zone drops steadily during
the AGB phase, depressing all reaction rates.  The final
$^{26}$Al/$^{27}$Al value calculated for a given $T_P$ is thus
drastically altered by the choice of reference state used to provide
the radiative-zone conditions.  That significant $^{26}$Al production
would occur is not an issue.  Rather, the question is how does one
pick an appropriate combination of stellar reference state and $T_P$
for $^{26}$Al production.  Fig. \ref{fig:deltat} shows a graph of
steady-state $^{26}$Al/$^{27}$Al ratios in the envelope as a function
of $\log(T_P/T_H)$ for the different stellar reference states
considered.  It can be seen that there is a congruent value of $T_P$
for each reference state that will produce essentially the same
(within a factor of 3) $^{26}$Al/$^{27}$Al if the value of $T_P$ is
scaled with $T_H$ as the star evolves.  It follows that the $T_P$
required for $^{26}$Al production is within this framework
well-constrained and that a CBP model covering all the isotopes of C,
N, O, Mg, and Al can be reliably calculated for a given reference
state.  The particular value of $T_P$ is simply congruent to some
effective value of $T_P/T_H$ and is not otherwise fixed.  This is in
accord with the approach taken by Boothroyd \& Sackmann (1999) and
Messenger (2000) for the RGB, but with the congruence made evident by
$^{26}$Al production rates.

\section{Experimental consequences}
\label{sec:consequences}

We have constructed above a model that should be a good description of
AGB stars undergoing cool bottom processing, using only two
parameters.  We now examine the consequences of the rules established
by this model for observed isotopic abundances.

\subsection{Circumstellar grains}

\subsubsection{Regime of C/O$<1$}
\label{sec:oxides}

\begin{figure}
\plotone{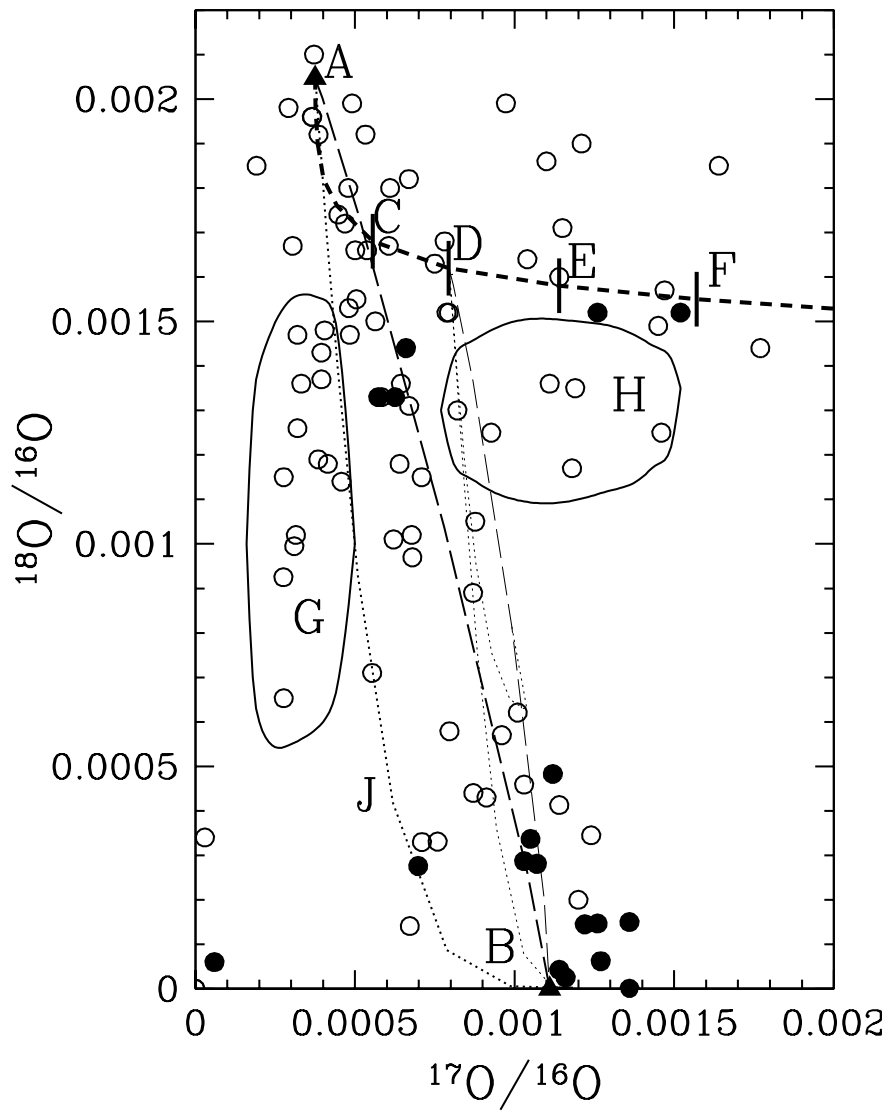}
\caption{Final envelope $^{18}$O/$^{16}$O and $^{17}$O/$^{16}$O with
oxide-grain data (summarized in Choi et al. 1998) superimposed.  The
dotted curve AJB indicates the leftmost limit of
compositions available starting from solar initial composition without
dredge-up.  Region G is therefore not accessible by CBP for a star
with solar initial composition.}
\label{fig:envelope17}
\end{figure}

Fig. \ref{fig:envelope17} exhibits the relationships between
$^{18}$O/$^{16}$O and $^{17}$O/$^{16}$O in the envelope after first
dredge-up (curve AF) and after CBP (thin dashed and dotted curves).
The available data on oxide grains are also shown. A substantial
population of the oxide grains can be seen to lie generally close to
the curve AF for AGB evolution after first dredge-up for stars of
varying mass, starting from initial solar isotopic composition.  (Data
are those summarized in Choi et al. 1998, and come from Hutcheon et
al. 1994, Huss et al. 1994, Nittler et al. 1994, Nittler et al. 1997,
Nittler \& Cowsik 1997 and Nittler et al. 1998; curve AF from
Boothroyd \& Sackmann 1999). Some points lie above the curve AF and
are plausibly explained by normal AGB evolution without CBP from
initial states with elevated $^{18}$O/$^{16}$O (Choi et al. 1998).
The trajectories for CBP of $^{18}$O/$^{16}$O and $^{17}$O/$^{16}$O
for a 1.5\ms\ star of solar composition are given by the line from
point D to point B, and the bounding curve for high \mdot\ following
Fig. \ref{fig:o17endpoints}a. It is seen that there is a population of
data that lies on this trend and a concentration of data approaching
and near to the equilibrium value at point B.  Following the rules
shown in Fig. \ref{fig:o17endpoints}b we see that data in the region
labelled H may be explained by stars of up to 1.8\ms\ initial
mass. They might equally be explained by an initial state of somewhat
lower $^{18}$O/$^{16}$O without CBP.  Lower-mass stars with solar
oxygen ratios directly modified by CBP would yield grains which lie in
the region bounded to the left by the curve AJB and the line AB,
corresponding to a range in \mdot\ and $T_P$.  However, the data in
region G can not be explained by AGB evolution from solar initial
composition, with or without CBP; these must involve sources with very
low to supersolar initial $^{18}$O/$^{16}$O values and subsolar
$^{17}$O/$^{16}$O values -- presumably representing different stages
of Galactic evolution.  The issue of how the abundances of the rare
isotopes $^{17}$O and $^{18}$O evolve over time in local material and
in an average over the Galaxy is complex (see Timmes \& Clayton 1996;
Nittler \& Cowsik 1996; Alexander \& Nittler 1999).  It is evident
that the wide variety of oxygen isotopic abundances found in
circumstellar oxides preserved in meteorites requires a diverse set of
stellar sources with quite different initial isotopic abundances and
metallicities.  Inspection of Figure \ref{fig:co} shows that for C/O
$< 1$ in the stellar envelopes, there is a wide range of $T_P$ that
produces the oxygen compositions shown in Figure \ref{fig:envelope17},
allowing a plausible explanation of a wide range of
$^{26}$Al/$^{27}$Al abundances.  The production of $^{26}$Al is quite
adequate to obtain values of $^{26}$Al/$^{27}$Al up to $\sim 6\times
10^{-2}$, the governing parameter being $7.68 \lesssim \log T_P
\lesssim 7.73$.  The values of $^{26}$Al/$^{27}$Al$\la 10^{-3}$ cannot
be achieved within the framework of standard AGB evolution (BGW99)
unless there are cases in which essentially negligible $^{26}$Al
production in the H shell is possible and the stellar lifetime is
still short enough to contribute to the solar-system grains.

\subsubsection{Regime of C/O$ > 1$}
\label{sec:carbides}

\begin{figure}
\epsscale{1.0}
\plottwo{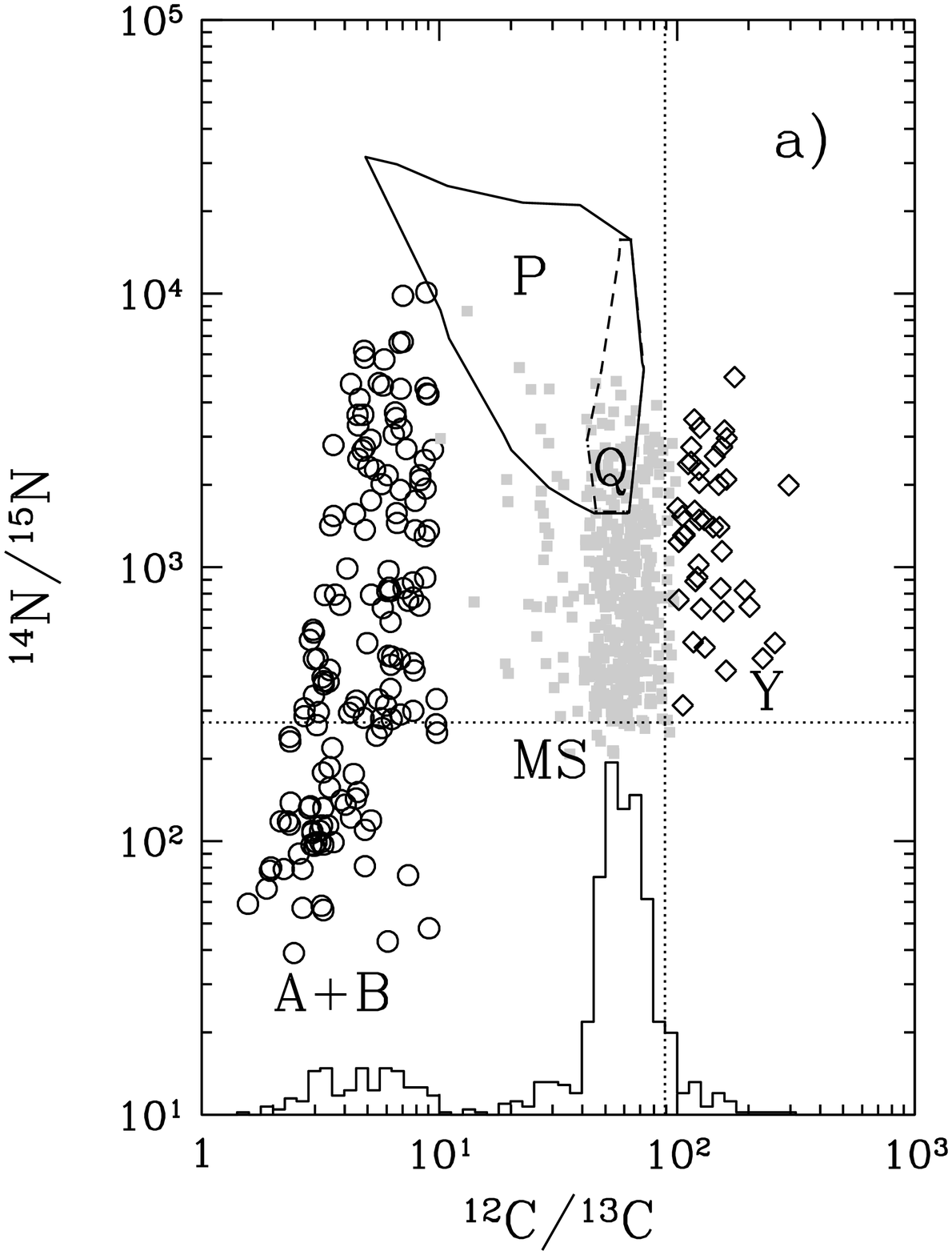}{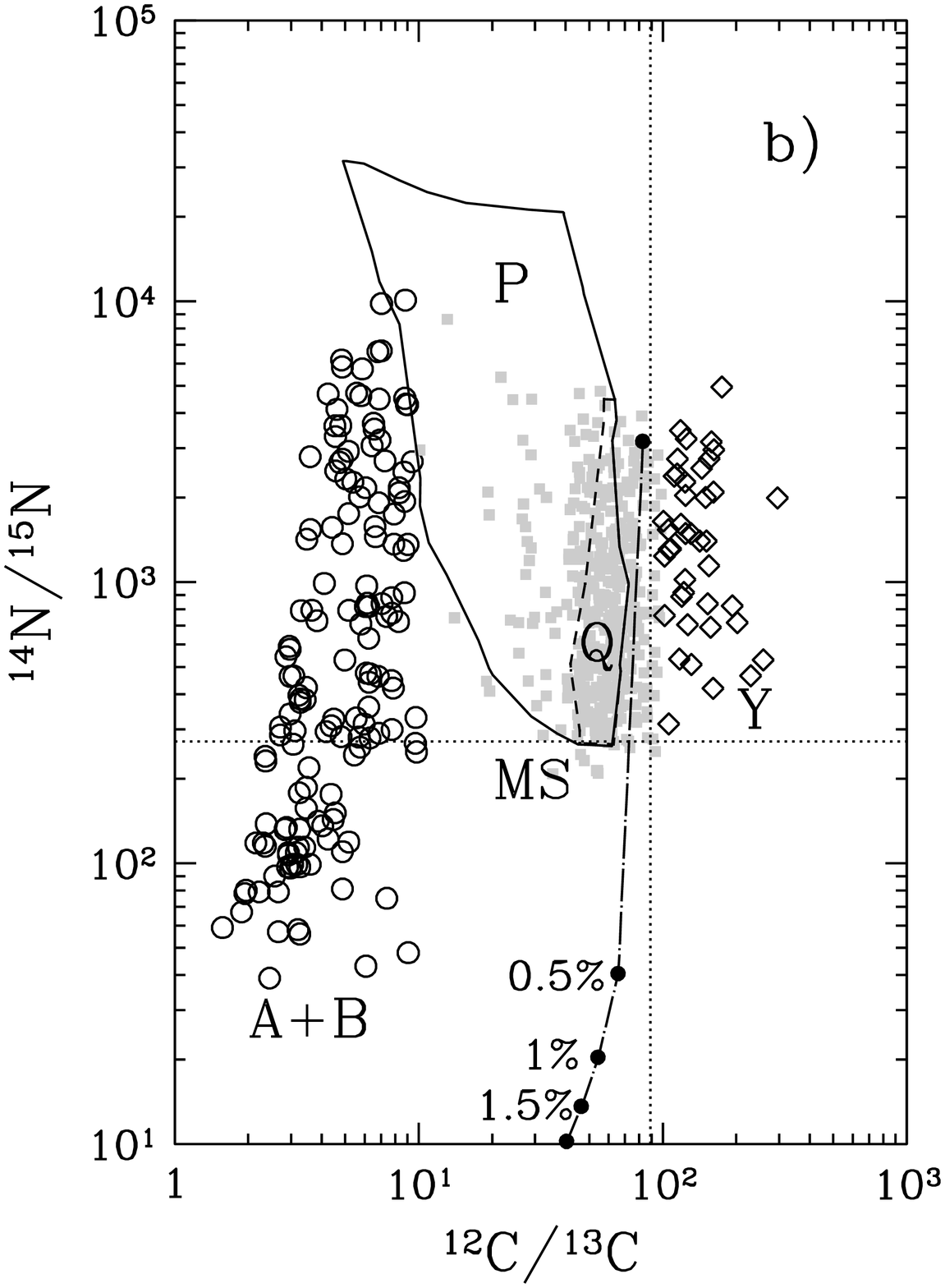}
\epsscale{0.7}
\caption{$^{12}$C/$^{13}$C and $^{14}$N/$^{15}$N of SiC grains
compiled by Amari et al. (2001b). Symbols are: $\circ$ A+B grains,
$\square$ mainstream (MS) grains, $\diamond$ Y grains.  The histogram
at the bottom is the frequency of each type (arbitrary linear units).
a) Regions P and Q contain all compositions with C/O$>1$ produced
during TDU and CBP; Q is with $^{26}$Al/$^{27}$Al$> 10^{-3}$.  b) same
model as (a), but for $^{14}$N/$^{15}$N $=270$ when the star arrived
on the AGB, corresponding to an intial $^{14}$N/$^{15}$N $\sim 70$
when the star formed.  The dot-dashed curve shows compositions that
result when a typical low-$^{15}$N/$^{14}$N composition from an AGB
star mixes with ejecta from a nova outburst.  Points along the curve
indicate the fraction of the composition made up of nova ejecta (nova
composition from Starrfield et al. 2001)}
\label{fig:amari1}
\end{figure}

The regime where carbon or carbide grains may form under equilibrium
conditions requires C/O $> 1$. There are abundant data on
$^{12}$C/$^{13}$C and $^{26}$Al/$^{27}$Al in circumstellar carbides
recovered from meteorites (e.g., Amari, Lewis \& Anders 1994; Hoppe et
al. 1994; Hoppe \& Ott 1997; Huss, Hutcheon \& Wasserburg 1997).  Most
recently the results have been summarized and discussed by Amari et
al. (2001a,b).  In addition to the $^{12}$C/$^{13}$C and
$^{26}$Al/$^{27}$Al data, there are results on $^{14}$N/$^{15}$N as
well as on Ti and Si isotopes.  In the following discussion, we will
focus on the so-called ``mainstream'' (MS) grains, ``A and B'' grains
and ``Y'' grains. These classifications are dominantly based on the
ratios $^{12}$C/$^{13}$C and $^{14}$N/$^{15}$N (see Amari et al. 2001b
for more extensive discussion).  The mainstream grains are
overwhelmingly the most abundant of the circumstellar SiC grains
found. Figure \ref{fig:amari1}a shows the distribution of grains in
the $^{12}$C/$^{13}$C, $^{14}$N/$^{15}$N diagram.  We first consider
the mainstream grains. As noted by many previous workers, the
$^{12}$C/$^{13}$C of mainstream (MS) grains lies in the region
expected for the envelopes of AGB stars of approximately solar
metallicity. The spread of $^{12}$C/$^{13}$C extends from slightly
above the solar value of 89 down to 20, with most grains between 40
and 80. We also show the whole region (labelled P and Q) that is
accessible by CBP in the numerical model with C/O $> 1$.

\begin{figure}
\plotone{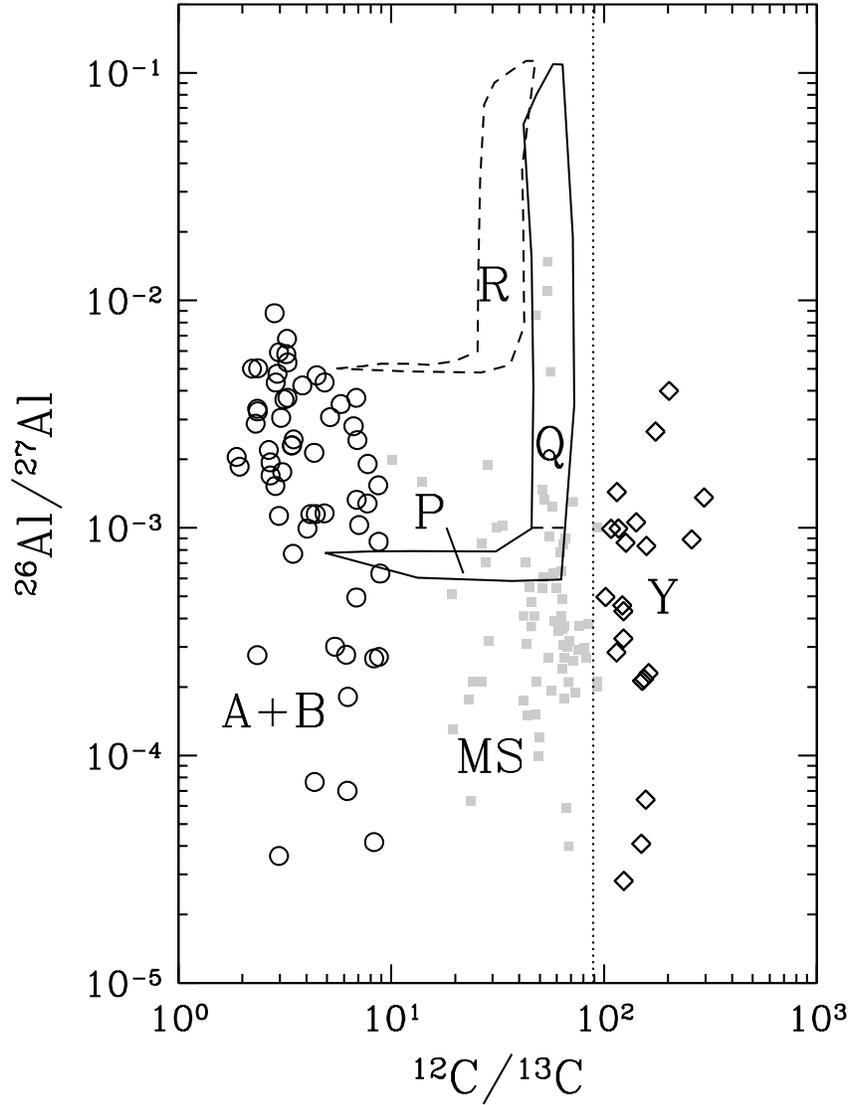}
\caption{C and Al isotopic compositions of SiC grains compiled in
Amari et al. (2001b; symbols as in Fig. \protect\ref{fig:amari1}).
The region P-Q contains all compositions produced with TDU and CBP.
For all values of $T_P$ and \mdot\ considered (except for the small
part of region P with $^{12}$C/$^{13}$C$\ga 50$, these are the same
regions as in Fig. \protect\ref{fig:amari1}).  Region R is analogous
to P-Q, but for models which assume that extensive CBP produced
$^{26}$Al/$^{27}$Al$= 10^{-2}$ and $^{12}$C/$^{13}$C=3 before TDU, and
then continued throughout TDU.}
\label{fig:amari3}
\end{figure}

The observed $^{26}$Al/$^{27}$Al versus $^{12}$C/$^{13}$C is shown in
Fig. \ref{fig:amari3}.  The horizontal segment P with low
$^{12}$C/$^{13}$C corresponds to the high-$^{14}$N/$^{15}$N region
also labelled P in Fig. \ref{fig:amari1}a.  This is also the region in
Fig. \ref{fig:13vCOtdu} of low $T_P$ and C/O$>1$, running from the
solid triangle (no CBP) down to $^{12}$C/$^{13}$C$\sim 5$ at $\dot{M}
> 10^{-5.6}\ms$/yr.  There is no $^{26}$Al produced as $T_P$ is low
along this segment of the curve.  The value of $^{26}$Al/$^{27}$Al
$\sim 8\times 10^{-4}$ in region P of Fig. \ref{fig:amari3}
corresponds to the total $^{26}$Al produced in the conventional
treatment of third dredge-up (BGW99; see Fig. \ref{fig:integral}a).
If we include an earlier episode of CBP (prior to TDU) sufficient to
give $^{26}$Al/$^{27}$Al $\sim 5\times 10^{-3}$, then the envelope
that is accessible for C/O $> 1$ is region R of Fig. \ref{fig:amari3}.
The observed $^{26}$Al/$^{27}$Al for the range of $^{12}$C/$^{13}$C
found in the MS grains is well represented by CBP.  However, the
samples with $^{26}$Al/$^{27}$Al $< 6\times 10^{-4}$ cannot be
explained by the model (independent of C/O).  A substantial fraction
of these grains (and of oxide grains) thus has less $^{26}$Al than
predicted for conventional AGB evolution in stars of $\sim 1
M_{\odot}$ (BGW99; Mowlavi \& Meynet 2000).  It is possible that this
reflects the normal AGB evolution at somewhat lower stellar masses, or
that there is some variability in the non-CBP $^{26}$Al production in
AGB stars as discussed in Sec. \ref{sec:intro}.

A fundamental problem with the SiC grains is exhibited by the
$^{14}$N/$^{15}$N data shown in Fig. \ref{fig:amari1}a. Here the
majority of MS grains lie below $^{14}$N/$^{15}$N $= 10^3$, and the
lower values are near the solar value.  This is exceedingly difficult
to explain with AGB stars in any evolution model that assumes solar
abundances as the initial state.  First dredge-up will destroy roughly
half of the $^{15}$N in the envelope and double the $^{14}$N, so that
the resulting value of $^{14}$N/$^{15}$N should be $\sim 10^3$.  Then
any CBP will only destroy more $^{15}$N and produce more $^{14}$N,
increasing the ratio.  To explain the high $^{15}$N, it has been
proposed by Huss, Hutcheon \& Wasserburg (1997) that the $^{18}{\rm
O}(p,\alpha)^{15}$N reaction rates are far greater than the accepted
values. If this is the case, then all of the $^{15}$N calculations
presented here are in error.  The nuclear data suffer from
poorly-known parameters for a low-energy resonance, but the most
recent work (Champagne \& Pitt 1986) contradicts earlier indications
of an additional large subthreshold resonance that would produce the
much higher rate advocated by Huss et al. (1997).  If we assume that
the calculations presented here are correct and that the MS grains
represent RGB and AGB evolution, then it is necessary to provide an
alternative hypothesis.  If the MS grains with $^{14}$N/$^{15}$N $<
10^3$ were the products of evolution on the RGB prior to first
dredge-up then to produce grains with no change from the initial N
composition, this would require RGB stars that have C/O $> 1$ and have
done nothing to their N isotopes.  No evidence for such stars, either
theoretical or observational, has been found.  If we consider that the
stars which produced the grains with low $^{14}$N/$^{15}$N were the
products of normal first dredge-up (DU) and subsequent AGB evolution,
then it is necessary that the stars' intial composition have
$^{14}$N/$^{15}$N far less than the solar value.  In all cases, first
DU increases $^{14}$N/$^{15}$N by a factor of $\sim 3.7$.  If we
assume that the typical initial state had $^{14}$N/$^{15}$N $\sim 70$,
then first DU would give the solar value and subsequent CBP would
essentially fill the region of the mainstream grains.

Following these considerations, we carried out calculations in which
we increased the assumed initial $^{15}$N inventory so that
$^{14}$N/$^{15}$N $= 270$ (the solar value) when the star arrives on
the AGB.  We obtained results for all the pertinent isotopes with this
initial composition in the numerical model.  The results for O, C, and
Al are of course not affected.  The results for $^{14}$N/$^{15}$N and
$^{12}$C/$^{13}$C are shown in Fig.  \ref{fig:amari1}b.  It can be
seen that the whole mainstream region (not just the upper part) is
covered with this shift in initial $^{14}$N/$^{15}$N.  We note that
adjusting the initial $^{14}$N/$^{15}$N by decreasing the amount of
$^{14}$N while holding all else fixed cannot provide the same
adjustment of the final $^{14}$N/$^{15}$N.  This is because first
dredge-up with initially solar abundance ratios among the metals
doubles the $^{14}$N mass fraction in the envelope.  Adjustments to
the initial $^{14}$N inventory do not affect the $^{14}$N gained at
first dredge-up, so they can reduce $^{14}$N/$^{15}$N on the AGB by no
more than a factor of two below the standard models.  The SiC grains
require a reduction by almost a factor of 4.

It is not clear that the solar $^{14}$N/$^{15}$N value is in any way
the proper choice for the ISM sampled in forming the stars which are
the source of the MS grains.  The data on oxide grains clearly
demonstrate that the stellar sources must come from different
molecular clouds which represent a wide range in degree of chemical
evolution.  The requirement that the bulk solar values represent a mix
of diverse sources is found in many examples.  This is particularly
the case for rare isotopes such as $^{13}$C, $^{15}$N, $^{17}$O, and
$^{18}$O.  We recall that $^{17}$O/$^{18}$O in the ISM (Penzias 1981)
has been found to be distincly different from the solar value (factor
of $1.7\pm 0.1$).  We consider that there has been evolution of these
isotopic abundances over Galactic history, that the abundances of some
isotopes are quite different today than at the time the solar nebula
formed, and that the abundances, particularly of very rare or
low-abundance isotopes will be distinctly different in different
places, even at the same time.  As is the case for oxygen, the
$^{14}$N/$^{15}$N values in the mainstream grains are clear
indications that the initial values for some sources cannot be the
solar value but must be greatly enhanced in $^{15}$N.  This conclusion
is unavoidable if the grains are produced by AGB stars.

Many grains assigned to the MS group exhibit clear $s$-process
signatures in several heavy elements which are attributed to
nucleosynthesis in AGB stars as noted by Hoppe \& Ott (1997) and
Gallino, Busso, \& Lugaro (1997).  The more abundant isotopic data on
Si and Ti in SiC grains show that these elements are dominated by
effects associated with galactic evolution (Gallino et al. 1994) with
small modifications attributed to $s$-processing.  The Si and Ti data
are not useful for our arguments because they do not strongly reflect
the s process.  However, the important new generation of measurements
on individual mainstream SiC grains for many elements show clear
$s$-process signatures in several isotopes of different elements
(Nicolussi et al. 1997; Nicolussi et al. 1998; Pellin et al.  2000).
In summary, there are compelling reasons to assign MS grains to AGB
stars, as recognized by many workers.  The largest problem with this
assignment appears to be the $^{14}$N/$^{15}$N data.  We conclude that
this can most simply be resolved by assuming that the initial
$^{14}$N/$^{15}$N typically available in molecular clouds $\sim
4.6\times 10^9$ yr ago must have ranged from $^{14}$N/$^{15}$N $\sim
70$ upward.  The ``solar'' value is considered to be the result of a
sampling of material that had previously undergone some prior
processing in AGB stars.

The N isotopic compositions determined for the ISM seem to support the
idea that $^{14}$N/$^{15}$N increases from low values in regions of
low astration to higher values in regions of high astration.  Chin et
al. 1999 find $^{14}{\rm N}/^{15}{\rm N} \sim 100$ for the Large
Magellanic Cloud, while higher values of 270 in the solar system,
200--600 in the Galactic disk, and $> 600$ at the Galactic center are
found in places with presumably increasing astration (see Wilson \&
Rood 1994 and references therein). It is thus plausible that
$^{14}$N/$^{15}$N was typically much lower at earlier times and
certain that it is variable.  On the other hand, it is generally
claimed (e.g. Wilson \& Rood 1994) that there is a gradient of
$^{14}$N/$^{15}$N with galactocentric distance in the Galactic disk,
in a direction suggesting that $^{14}$N/$^{15}$N decreases with
astration.

Inspection of Figs. \ref{fig:amari1}a,b and \ref{fig:amari3} shows the
data on Y grains, which are classified as distinctive almost solely
because their $^{12}$C/$^{13}$C is greater than the solar value. The
distributions of these grains in both $^{14}$N/$^{15}$N
vs. $^{12}$C/$^{13}$C and $^{26}$Al/$^{27}$Al vs.  $^{12}$C/$^{13}$C
are otherwise the same as those for the MS grains.  If one considers
$^{12}$C dredge-up in models of AGB stars, it is evident that an
increase in the amount of dredge-up by a factor of 2 or 3 for a small
proportion of AGB stars would yield a carbon star with Y-grain-like
composition.  In this case, the Y grains are not distinguishable from
the more general population of grains produced by AGB stars but
reflect some variation in C dredge-up.  Our preferred interpretation
is that they simply reflect the high-end tail of the C/O distribution
resulting from dredge-up, which may or may not result from variation
of metallicity (i.e., variation of intial O).  The extent to which the
variability of $^{12}$C dredge-up is possible within the framework of
AGB models remains to be tested.  Other proposals to explain the Y
grains are based on the dependence of dredge-up on metallicity, as
argued by, e.g., Amari et al. (2001b) and references therein on the
basis of anomalies in the Si isotopic composition.

We now consider the A and B grains.  As discussed by Amari et al.
(2001b), these grains are defined by low $^{12}$C/$^{13}$C$< 10$.
They are also found to exhibit $^{14}$N/$^{15}$N ranging over more
than two orders of magnitude and extending to below the solar value.
We consider them here as representing a group of grains produced by a
generic process.  It is evident that AGB stars can produce
$^{12}$C/$^{13}$C as low as 4 at low $T_P$ but they can not produce a
wide range in $^{14}$N/$^{15}$N and $^{26}$Al/$^{27}$Al and
simultaneously have $^{12}$C/$^{13}$C $\lesssim 12$ and C/O $> 1$.  It
follows that an AGB source at approximately solar metallicity can not
be the origin of such grains.  Further, we do not consider that this
condition can be directly related to low metallicity.  It should also
be noted that a considerable number of these grains have
$^{12}$C/$^{13}$C $< 3$, ranging down to $^{12}$C/$^{13}$C $= 1.57 \pm
0.06$ and $1.8 \pm 0.01$ (Amari et al., 2001b).  From the existing
knowledge of reaction rates, values below 3 can not be achieved on
either the RGB or the AGB with or without CBP.  It follows that either
there is some significant ($\times 2$) error in the reaction rates or
that these grains were produced under conditions quite different than
the CNO burning outlined in Section \ref{sec:carbon} and as discussed
by other workers.  A considerable number of the grains have
$^{14}$N/$^{15}$N extending down to $^{14}$N/$^{15}$N $= 40$, far
below the solar value.  It was suggested by Amari et al. (2001b) that
this may be produced by hot H-burning (far above AGB shell-burning
temperatures), possibly associated with novae.  While it is reported
that many CO novae condense carbon dust, it should be noted that only
three of the 31 recent novae listed by Gehrz et al. (1998) have been
observed to have C/O $> 1$ .  The formation of different types of
grains (e.g., carbides or silicates) in the same nova may reflect
sequential condensation (see Gehrz et al.).  The calculated yields for
thermonuclear runaways (TNR) in novae almost exclusively have C/O$ <
1$ (Starrfield et al. 2001, and S. Starrfield 2001, personal
communication). The most recent calculations (S. Starrfield 2001,
personal communication and Starrfield et al., in preparation) show
that C/O $< 1$ for white dwarf masses of 1.25 and 1.35 $M_{\odot}$.
An extremely high $^{17}$O/$^{16}$O ratio ($> 1$) is predicted in all
cases.  A search for implanted oxygen with such extreme $^{17}$O
enrichments in graphite or SiC grains would therefore be very
important.  It is known that extreme values of $^{12}$C/$^{13}$C and
$^{14}$N/$^{15}$N ($\lesssim 1.5$), as well as substantial $^{26}$Al
are predicted for novae yields (cf. Starrfield et al., 2001).
Nonetheless, it is difficult to connect the A and B grains as a group
to novae sources.  There are only two reported cases of high
$^{26}$Al/$^{27}$Al in SiC grains with extremely low $^{12}$C/$^{13}$C
and $^{14}$N/$^{15}$N.  They belong to the very small class (6 known)
of carbon-rich grains attributed to novae on the basis of C, N, and Si
isotopic compositions, and they are not A and B grains (Amari et
al. 2001c).  No models of novae give $^{14}$N/$^{15}$N far above
unity.  The yields of C, N and O isotopes are all very roughly
commensurate (factor-of-ten) so that the wide spread in
$^{14}$N/$^{15}$N cannot be achieved with low $^{12}$C/$^{13}$C.
Thus, the high $^{14}$N/$^{15}$N values found for some A and B grains
cannot be attributed to novae.  The A and B grains appear to suggest
some connection with novae; however, this cannot be due to the TNR on
ONeMg white dwarf models for which nucleosynthesis is usually
computed.  One possibility for a direct nova connection would be if
the fragile $^{15}$N could be destroyed during the later stages of the
explosion.  The models of novae so far considered have mostly high
accretion rates, and it is conceivable that low accretion rates which
do not initiate full TNR could produce partially processed material
with isotopic characteristics resembling A and B grains.  The true
novae represent extremely energetic bursts punctuating a slower,
ongoing accretion process.  If we assume that the A and B grains were
produced by a mixture of accreting AGB matter that is not processed in
TNR and a component that was produced through TNR, then we immediately
face the problem that the nova yield ratio $Y_{13}^N/Y_{15}^N$ is
typically $\sim 0.1$, while the AGB matter with which it is mixed has
almost no $^{15}$N.  A mixing trajectory between AGB compositions and
nova compositions would therefore move rapidly from the AGB values to
those with very low $^{14}$N/$^{15}$N, but at high $^{12}$C/$^{13}$C
(see Fig. \ref{fig:amari1}b).  Such a mixing curve would not pass
through the A and B grains.  Region P of Fig. \ref{fig:amari1}a,b
contains AGB compositions with extensive CBP that could mix with a
nova composition to produce a mixing curve passing through the A and B
grains, but only for a small corner of the parameter space where CBP
produces $^{12}$C/$^{13}$C$\sim 10$ and $^{14}$N/$^{15}$N$\sim 2\times
10^4$.  Since this composition corresponds to low $T_P$, any $^{26}$Al
found in the grains would have to come from the novae.  In summary, we
have no self-consistent mechanism to propose for A and B grains.  They
appear to reflect different mechanisms than are currently available in
stellar models.  If we assume the classification of A and B grains
into a single group to be invalid and consider instead subsets of the
A and B grains, the problem is not removed.

\subsection{Astronomical implications}

We now consider what evidence concerning cool bottom processing on the
AGB should be accessible to astronomical observations.  As in the case
of the pre-solar grains, the potential for converting large amounts of
C to N is crucial, and we divide up the results into those expected
for AGB stars with C/O$>1$ and with C/O$<1$.  This discussion applies
to stars that have been on the AGB long enough to have become C stars
without CBP; for the SEM(1.5) model, this corresponds to the star
exceeding a luminosity of $1.2\times 10^4L_\odot$.

\begin{figure}
\epsscale{1}
\plottwo{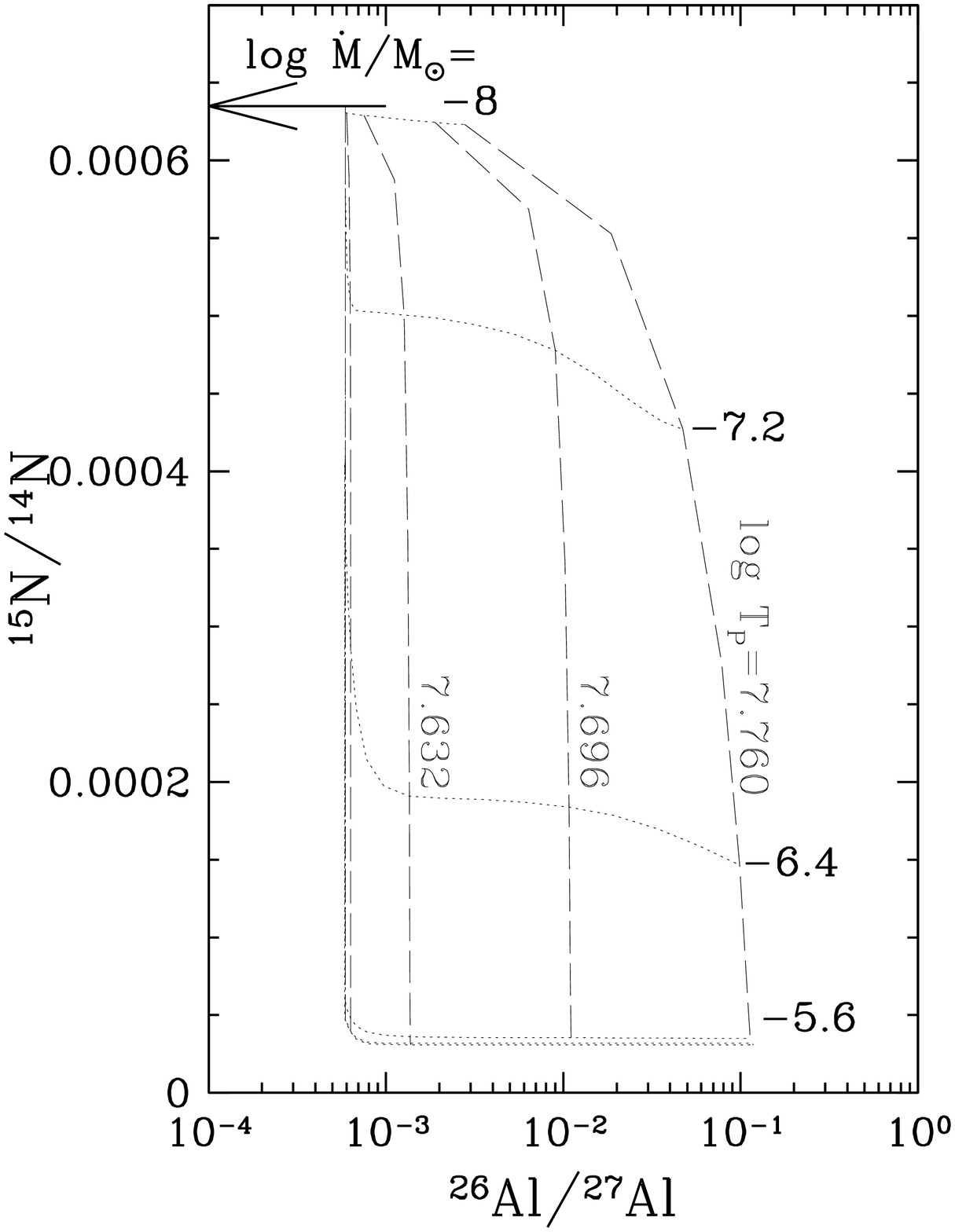}{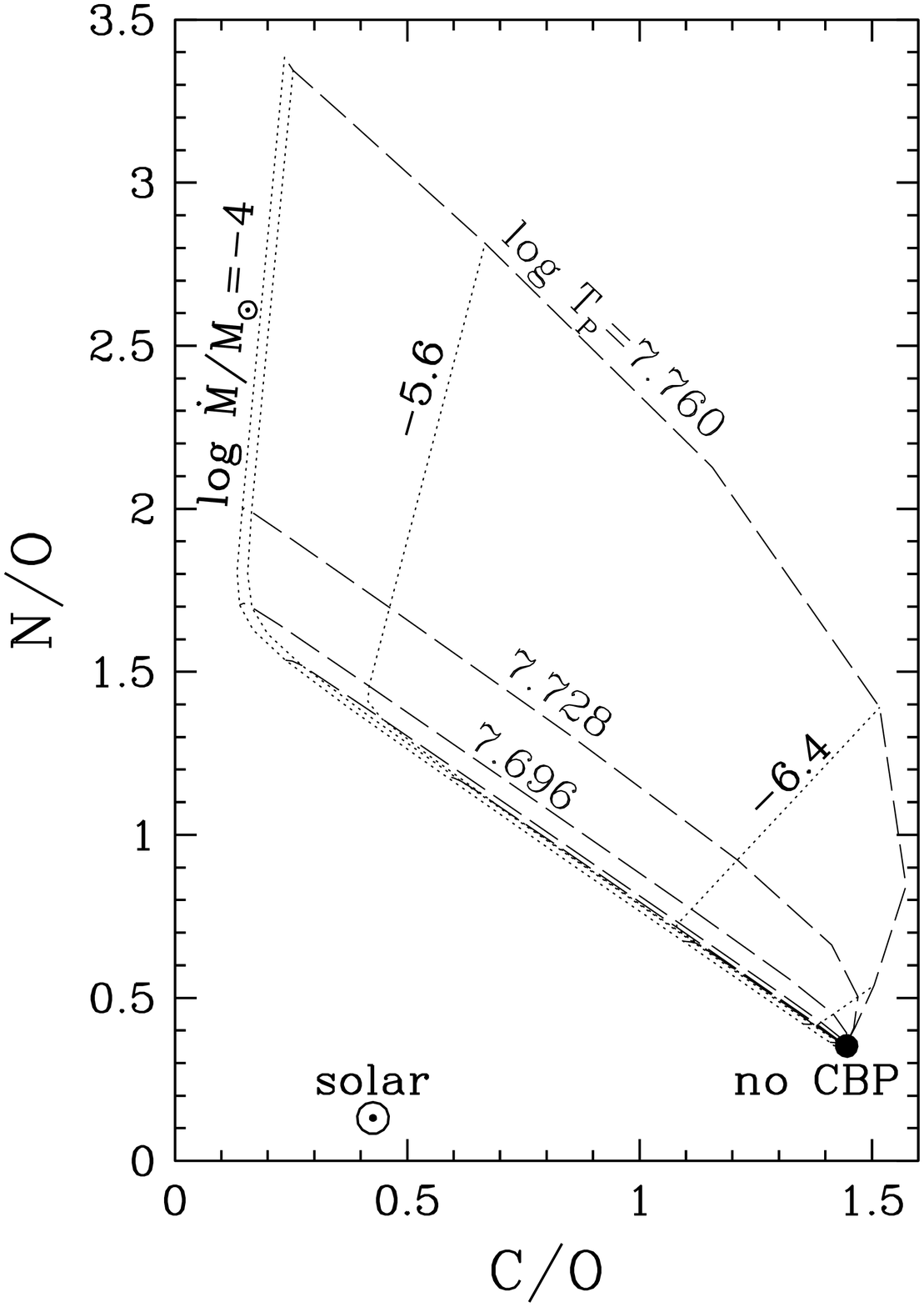}
\epsscale{.7}
\caption{Final envelope compositions for the numerical model. a)
$^{15}$N/$^{14}$N decreases from an initial value of $6.3\times
10^{-6}$ toward a final value $\sim 3\times 10^{-5}$ for all $T_P$,
but with an effectiveness that varies mainly with \mdot.  The arrow at
the top indicates the envelope composition after first dredge-up.  b)
final values of N/O and C/O.  At most temperatures, this reflects
varying degrees of conversion of C into N; the two curves with the
$\log T_P>7$ reflect additional destruction of O to make more C and N.
The point labelled ``solar'' indicates solar-system composition.}
\label{fig:n15}
\end{figure}

Fig. \ref{fig:co} shows that for almost any value of $\log T_P >
7.60$, $\mdot\ga 10^{-6}\ms$/yr can prevent carbon-star formation.
Low values of \mdot\ are therefore required if CBP occurs in C stars.
Low \mdot\ requires oxygen isotopes with little or no change from the
values set by first dredge-up; this means $^{18}$O/$^{16}$O$\sim
0.0015$ and wide variation of $^{17}$O/$^{16}$O with stellar mass
(along the thick-dashed curve of Fig. \ref{fig:o17endpoints}).  It may
further be expected that $10^{-4} <\ ^{15}\mathrm{N}/^{14}\mathrm{N} <
6\times 10^{-4}$ and that incomplete conversion of C to N in the
envelope produces $0.5 < \mathrm{N/O} < 0.8$
(cf. Figs. \ref{fig:n15}a,b) as well as $^{12}$C/$^{13}$C$>20$.  In
principle, it is possible to obtain a carbon star with
$^{12}$C/$^{13}$C$<20$ if $\log T_P \lesssim 7.6$ and \mdot\ is high,
but this case constitutes a very small part of the parameter space.
Any level of $^{26}$Al enrichment is compatible with a C-rich star,
depending on $T_P$.  We thus find that contrary to usual expectations
(e.g. Arnould and Prantzos 1999; Mowlavi \& Meynet 2000), normal
low-mass C stars with low \mdot\ could produce abundant $^{26}$Al and
they are reasonable targets for gamma-ray line observations by
high-energy satellite experiments like INTEGRAL.  However, low
$^{12}$C/$^{13}$C requires low $T_P$ that is not compatible with
significant $^{26}$Al/$^{27}$Al.

If C/O$<1$ in the envelope of an evolved TP-AGB star or planetary
nebula, it requires $\mdot \gtrsim 10^{-6}$\ms/yr.  All of the signs
of CBP that become more pronounced with greater \mdot\ should then be
present: $^{18}$O/$^{16}$O$\lesssim 8\times 10^{-4}$
(Fig. \ref{fig:o18endpoints}), $^{15}$N/$^{14}$N$\lesssim 2\times
10^{-4}$ (Fig. \ref{fig:n15}a), and $5
<\,^{12}\mathrm{C}/^{13}\mathrm{C} < 70$ (Fig. \ref{fig:13vCOtdu}).
The large quantity of carbon destroyed also requires N/O$\gtrsim 1.5$
(Fig. \ref{fig:n15}b).  All of this processing should simultaneously
destroy $^3$He.  Individual objects may be found with N/O$\geq 3$ if
$T_P$ is very high.  Again, since $^{26}$Al production is independent
of \mdot, it should occur equally often in O-rich as in C-rich
environments if CBP is active.

Large enrichments of N/O in planetary nebulae may be important signs
of cool bottom processing on the AGB.  High N/O is frequently observed
and is a defining characteristic of the Type I planetary nebulae
(cf. Corradi \& Schwarz 1995) that make up some 20--25\% of observed
planetary nebulae.  Past expectations have been that N enrichment
requires HBB (Peimbert \& Serrano 1980; van den Hoek \& Groenewegen
1997) and thus intermediate-mass stars.  CBP would permit N-rich
planetary nebulae to be made from low-mass ($< 3\ms$) stars.  A search
of planetary nebulae for other signs of CBP at high \mdot\ (those
associated with O-rich AGB stars), combined with good stellar mass
estimates, would indicate whether the N originated in CBP.
Measurements of $^{14}$N/$^{15}$N, N/O, $^{18}$O/$^{16}$O,
$^{17}$O/$^{16}$O, and $^{12}$C/$^{13}$C would provide clear tests of
CBP.  Evidence for moderate N enrichment in a single low-mass
planetary nebula has already been found by P\'equignot et al. (2000).
It would be of great interest if more of these objects are found.
Since the same H-burning cycles that characterize CBP are also active
in HBB, it would be necessary to determine initial masses of the
observed stars and nebulae in order to distinguish CBP and HBB.

\subsection{Galactic inventories}

The extensive processing of stellar envelopes discussed above may have
consequences for the overall Galactic inventories of at least two
nuclides: $^3$He and $^{26}$Al.  Observations of $^3$He are difficult,
and have so far been confined to the local ISM and some planetary
nebulae (Gloeckler \& Geiss 1998; Balser et al. 1998; Bania, Rood \&
Balser 2002).  The results indicate that the Galaxy contains no more
$^3$He than one would expect from its primordial inventory plus the
result of pre-main-sequence deuterium burning.  It is presently not
understood why so little $^3$He has apparently been made over the
lifetime of the Galaxy.  Low- and intermediate-mass stars should
produce large amounts of $^3$He during their main-sequence evolution
(e.g., Sackmann \& Boothroyd 1999).  First dredge-up should bring much
of this $^3$He into the stellar envelopes, where subsequent mass loss
allows its escape into the interstellar medium.  Conventional stellar
evolution provides no way to destroy this dredged-up $^3$He.  It has
been noted in earlier work (WBS; Charbonnel 1995) that cool bottom
processing on the red giant branch could provide a path to destroy
$^3$He, using the reactions $^3{\rm He}(^3{\rm He},2p)^4{\rm He}$ and
$^3{\rm He}(\alpha,\gamma)^7{\rm Be}$.  When we include these
reactions in our nuclear-reaction network, we find that extensive
$^3$He destruction also occurs in the AGB cool bottom processing
scenario presented here.  The results of CBP for $^3$He are directly
analogous to those for $^{18}$O: there is a threshold temperature
above which destruction can become rapid.  We find a threshold
temperature for thorough $^3$He destruction of $\log T_P \sim 7.65$;
factor-of-ten depletion may be reached for $\log T_P=7.60$.  Just as
for $^{18}$O, the final abundance of $^3$He in the stellar envelope
then depends on the rate and duration of mixing, with duration limited
by the stellar lifetime.  For both $^3$He and $^{18}$O, essentially
complete (more than a factor of 10) destruction occurs at $\mdot\ga
3\times 10^{-6}\ms$/yr as long as the threshold temperature is
exceeded.  As a result, low-mass stars that undergo cool bottom
processing on the AGB may not be net producers of $^3$He over their
lifetimes, and this may help explain the apparently small amount of
$^3$He production that has occurred.

An observational test of whether $^3$He is destroyed in this way may
be obtained by spectroscopic measurements of $^{18}$O/$^{16}$O in CO
lines from AGB stars or planetary nebulae, since both $^{18}$O and
$^3$He destruction should occur simultaneously.  The threshold $T_P$
for $^{18}$O destruction is lower than that for $^3$He destruction, so
$^{18}$O destruction is required if $^3$He is destroyed.  The
connection between $^3$He and $^{18}$O destruction is also important
for another reason: $^3$He destruction by CBP must occur in
essentially all low-mass stars if CBP is to account for the lack of
evolution of the $^3$He inventory.  This implies that essentially all
low-mass stars should destroy most of the $^{18}$O in their envelopes,
and in turn that the preponderance of oxide grains originating in
these stars should have $^{18}$O/$^{16}$O $< 10^{-4}$.  Examination of
Fig. \ref{fig:envelope17} shows that a large fraction of the sources
sampled by the oxide grains have higher values, suggesting that there
might not be sufficient CBP to destroy the extra $^3$He produced in
low-mass stars.  This problem requires further attention.

The Galactic inventory of $^{26}$Al is also of interest.  The Galaxy
contains $\sim 3\ms$ of $^{26}$Al, as inferred from observations of
the gamma-ray line from its decay (Mahoney et al. 1984; Kn\"odlseder
et al. 1996).  Since $^{26}$Al has a mean lifetime of 1 Myr, this
requires ongoing nucleosynthesis.  The sites of this nucleosynthesis
remain uncertain (e.g., Kn\"odlseder et al. 1999).  If the sources of
the $^{26}$Al are AGB stars (which neither create nor destroy
$^{27}$Al) and if the rate of injection of $^{26}$Al into the ISM
matches the rate of decay, then the material being ejected into the
ISM must have $^{26}$Al/$^{27}$Al satisfying
\begin{equation}
\frac{^{26}{\rm Al}}{^{27}{\rm Al}}
  \simeq  \frac{M_G /\tau_{26}}{X_{27}M_* R},
\end{equation}
where $M_G \sim 3 \ms$ is the Galactic $^{26}$Al inventory, $X$'s are
mass fractions, $M_*$ is the mass of ejecta from a single AGB star,
and $R$ is the death rate (stars per year) of AGB stars.  If $X_{27}$
is solar, then
\begin{equation}
\frac{^{26}{\rm Al}}{^{27}{\rm Al}}  =
\frac{5.4\times 10^{-2}\ms/{\rm yr}}{M_*R}
\label{eqn:inventory}
\end{equation}
in the ejecta (see Dearborn, Lee \& Wasserburg 1988).  It is quite
difficult to estimate $M_*R$.  Consider two cases, with stellar death
rates based first on the local density of giant stars and then on an
estimate of the local white-dwarf birth rate.  In either case, we must
extrapolate local stellar populations into totals for the Galaxy;
since the $^{26}$Al gamma-ray line emission seems to be concentrated
in the Galactic disk, we assume that the Galaxy consists only of the
thin-disk component described in Binney \& Merrifield (1998), and we
assume that the dying stars (i.e., those shedding their envelopes into
the ISM and evolving into white dwarfs) make up the same fraction of
stars everywhere in the Galaxy as they do locally.  Using the local
density of red giant stars from Binney \& Merrifield (1998) and
assuming that each of these stars sheds its envelope after a lifetime
of $10^8$ years, we find $R=1.3\ {\rm yr}^{-1}$.  If these stars each
eject $M_*\sim 1\ms$, then the ejecta must have
$^{26}$Al/$^{27}$Al$=4\times 10^{-2}$, from Eqn. \ref{eqn:inventory}.
Fig. \ref{fig:integral}a shows that this level could be produced by
CBP with $\log T_P\sim 7.75$.  Extensive CBP at high $T_P$ might
permit AGB stars to be a significant source of the Galactic $^{26}$Al.
However, the preponderance of circumstellar grains appear to be
produced in environments where $^{26}$Al/$^{27}$Al is much below
$4\times10^{-2}$.  If we use the white-dwarf birth rate of Drilling \&
Sch\"onberner (1985) and we extrapolate the local rate of $(2\ {\rm
to}\ 3) \times 10^{-13}\ {\rm pc}^{-1}\ {\rm yr}^{-1}$ to a total
Galactic rate, we obtain $R= 0.04$ to $0.06\ {\rm yr}^{-1}$.  This is
far too low to contribute to the Galactic inventory.  We conclude that
it is unlikely that a significant fraction of the $^{26}$Al observed
in gamma-ray line emission in the Galaxy was made in AGB stars.  This
conclusion has also been reached on the basis of the apparently clumpy
distrubution of $^{26}$Al in the Galaxy (Diehl and Timmes 1998).
However, cool bottom processing in AGB stars may provide sufficient
$^{26}$Al to permit detection in selected AGB stars with very low
$^{18}$O/$^{16}$O and high N/O.

\section{Conclusion}

The work presented in this paper demonstrates that extensive
predictions of the possible range of envelope compositions and
isotopic abundances of AGB stars may be made in a model of cool bottom
processing parametrized only by a mass-flow rate \mdot\ and a maximum
temperature $T_P$ using a simple calculation.

This calculation uses results from a stellar evolution model
calculated without CBP to provide the basic conditions needed to
determine the results of additional processing by extra mixing.  It
appears that significant composition changes have negligible feedback
on the stellar luminosity.  As a result, the post-processing approach
adopted here -- in which CBP is calculated separately from stellar
evolution -- appears to be reasonable.  Post-processing calculations
allow the range of possible effects of extra mixing to be determined
easily by exploring the space of $T_P$ and \mdot\ values once stellar
structures have been determined for an arbitrary star.  The results
reported in this work are based on a grid of 231 pairs of the two
parameters.  This covers the entire range that is both plausible and
potentially interesting, using an unsophisticated code and about one
day of processing time on a personal computer with 930 MHz clock
speed.  Moreover, given only the stellar structure and reaction rates,
one can obtain almost quantitatively correct analytical approximations
to these results as functions of \mdot\ and $T_P$ for C/O
(Eqn. \ref{eqn:cocondition} and Fig. \ref{fig:co}),
$^{26}$Al/$^{27}$Al (Eqn. \ref{eqn:prodrate-1} and
Fig. \ref{fig:integral}a), $^{18}$O/$^{16}$O (Eqn. \ref{eqn:decay} and
Fig. \ref{fig:mdotvs18}), and $^{17}$O/$^{16}$O (Eqn. \ref{eqn:mixing}
and Fig. \ref{fig:mdotvs17}).  Given the uncertainties concerning the
exact mass, evolutionary state, mass-loss rate, and extra-mixing
parameters corresponding to the star that produced an observed
presolar grain or planetary nebula, the analytic approximations are
reasonably accurate and display in a clear way the rules that govern
the outcome of CBP.

We found also that given the uncertainties (particularly the unknown
mixing mechanism, which will set the mixing depth), the detailed
evolution of the stellar structure below the stellar envelope (e.g.,
the radiative zone) is relatively unimportant.  The results from a
complete stellar evolution code without extra mixing are necessary to
specify the timescales, the evolution of envelope mass, the action of
dredge-up, and a representative structure for the region in which
nuclear processing takes place in the CBP model.  However, the
detailed evolution of this region need not be followed to obtain
useful results.  The major effect of the evolution of this region is
to change the $T_P$ that corresponds to a given amount of processing;
holding both the structure of this region and $T_P$ fixed is
equivalent to holding $T_P/T_H$ fixed in a model that follows the
detailed evolution, provided that both follow evolution of the
envelope in the same way.

The effects of CBP on stars of different masses or metallicities
(particularly low metallicity) can readily be determined once a
conventional non-CBP stellar model and radiative-zone structure are
obtained from conventional codes.

\acknowledgments

We gratefully acknowledge useful the use of the FRANEC code, granted
by O. Straniero, A. Chieffi and M.  Limongi, as well as extensive
comments and suggestions from J. Lattanzio and an anonymous referee.
Some incisive comments by R. Gallino were valuable and helpful.
M.B. wishes to thank the Lunatic Asylum and Caltech for hospitality.
This work was supported by NASA grant NAG5-11725 and by the Italian
MURST contract Cofin2000.  Caltech division contribution 8769(1080).

\appendix

\section{Reaction Rate Uncertainties}
\label{sec:rates}

Most of the reaction rates used in this paper are known to within
5--40\%, according to the NACRE evaluation (Angulo et al. 1999).
There are a few cases of larger uncertainty, which we now examine.  In
cases where more or less reliable error estimates exist, the NACRE
upper and lower limits to the reaction rates are often found by
multiplying by two the size of the statistical error bars one derives
directly from the experimental reports.  We consider the rates at a
temperature of $5\times 10^7\ {\rm K}$ as representative of the
crucial temperature range.

First, the NACRE compilation quotes uncertainties of a factor of two
for the rate of $^{18}{\rm O}(p,\alpha)^{15}{\rm N}$.  This affects
only the rate of $^{18}$O destruction at low $T_P$ ($\la 10^{7.62}$
K).  Above this temperature, increasing the destruction rate changes
nothing because even with the recommended rates, a single pass through
the radiative zone suffices to greatly reduce the $^{18}$O abundance
in the circulating matter.  Likewise, decreasing this rate will have
no effect as long as the lower rate still suffices to destroy all
$^{18}$O in a single pass; the threshold $T_P$ for this condition is
also about $10^{7.62}$ K.  Below $\log T_P=7.62$, factor-of-two
changes in the $^{18}$O-destruction rate translate into factor-of-two
changes in the final $^{18}$O/$^{16}$O.

The most important uncertainties are those concerning reactions which
produce and destroy $^{26}$Al.  The NACRE compilation quotes upper and
lower limits for the rate of $^{26}{\rm Al}^g(p,\gamma)^{27}{\rm Si}$
which are a factor of 47 higher and a factor of 86 lower,
respectively, than the recommended rate.  It should be kept in mind
that the NACRE recommended rate at these temperatures was determined
mainly by setting resonance strengths to be 1/10 of upper limits
derived from the shell-model calculation of Champagne, Brown \& Sherr
(1993; these are considerably lower than the experimental upper
limits).  Changing from the NACRE recommended rate to the NACRE
upper-limit rate reduces the equilibrium $^{26}$Al/$^{25}$Mg in
circulating material from near 100 to around unity.  As a result, low
\mdot\ then allows this ratio to equilibrate in the circulating
material, so that the proportionality of $^{26}$Al production to
$\mdot^{-1}$ in Eqn. \ref{eqn:mdotscaling} breaks down, and the amount
of $^{26}$Al returning to the surface becomes both lower and
independent of \mdot.  Although the final $^{26}$Al/$^{27}$Al in the
envelope is smaller in this case, it still remains high enough to
account for all oxide grains with only a very slight increase in the
maximum $T_P$.

The rate of the $^{26}$Al production reaction, $^{25}{\rm
Mg}(p,\gamma)^{26}{\rm Al}^g$, is quoted with an upper limit 75\%
above the recommended rate and a lower limit 41\% below the
recommended rate.  The NACRE rate and uncertainties for this reaction
are based on resonance strengths extracted by Iliadis et al. (1996)
from pre-existing data using improved DWBA methods.  Except at $\mdot
\la 10^{-7}\ms$/yr, where the independence of the envelope $^{26}$Al
enrichment rate from \mdot\ breaks down, the result of varying the
$^{25}{\rm Mg}(p,\gamma)^{26}{\rm Al}$ rate within the NACRE limits is
to scale the final $^{26}$Al/$^{27}$Al downward by as much as 40\% or
upward by up to 75\%.

The only other large uncertainties (greater than a factor of two) in
the reaction rates included in our code are for the
$^{27}$Al-producing and -destroying reactions $^{26}{\rm
Mg}(p,\gamma)^{27}{\rm Al}$, $^{27}{\rm Al}(p,\alpha)^{24}{\rm Mg}$,
and $^{27}{\rm Al}(p,\gamma)^{28}{\rm Si}$.  The first of these
reactions has an upper limit in the NACRE compilation which is 15
times the recommended rate (derived mainly from experimental work by
Champagne et al. 1990) and a lower limit which is 1/27 the recommended
rate.  The upper and lower limits to the $(p,\alpha)$ destruction rate
differ from the recommended rate by factors of 10 and 5, respectively.
To examine the allowed effect on $^{26}$Al/$^{27}$Al due to $^{27}$Al
production and destruction (which is negligible when using the
recommended rates), we ran models with all of the $^{27}$Al-producing
reaction rates (including the $^{26}$Al-destroying $^{26}{\rm
Al}(p,\gamma)^{27}{\rm Si}(\beta^+\nu)^{27}{\rm Al})$ set to the NACRE
upper limits and the $^{27}$Al-destroying reaction rates set to their
NACRE lower limits.  This made virtually no difference to the final
$^{26}$Al/$^{27}$Al beyond the effects of $^{26}{\rm
Al}(p,\gamma)^{27}{\rm Si}$ on $^{26}$Al production described above.
The corresponding calculation with the limits switched to minimize the
final $^{27}$Al abundance showed no significant difference from that
using the recommended rates.

\epsscale{.7}

\end{document}